%% file: main.tex
\documentclass{article}
\usepackage{PRIMEarxiv}
\usepackage[utf8]{inputenc} % allow utf-8 input
\usepackage[T1]{fontenc}    % use 8-bit T1 fonts
\usepackage{hyperref}       % hyperlinks
\usepackage{url}            % simple URL typesetting
\usepackage{booktabs}       % professional-quality tables
\usepackage{amsfonts}       % blackboard math symbols
\usepackage{nicefrac}       % compact symbols for 1/2, etc.
\usepackage{microtype}      % microtypography
\usepackage{lipsum}
\usepackage{fancyhdr}       % header
\usepackage{graphicx}       % graphics
\graphicspath{{media/}}     % organize your images and other figures under media/ folder
\usepackage{natbib}
\usepackage{amsfonts,amsmath,amssymb}
\usepackage[table]{xcolor}

\usepackage{subfigure,bm,yfonts,cleveref,xcolor}

\title{Time-lapse data matching using a recurrent neural network approach}

\author{
  Abdullah Alali \\
  KAUST \\
  Saudi Arabia, Thuwal \\
  \texttt{abdullah.alali.1@kaust.edu.sa} \\
  \And
  Vladimir Kazei \\
  Aramco Americas \\
  United States, Houston\\
  \texttt{vkazei@gmail.com } \\
  \AND 
\hspace{45pt} Bingbing Sun \\
\hspace{45pt}  Saudi Aramco \\
\hspace{45pt}  Saudi Arabia, Dhahran \\
\hspace{45pt}  \texttt{bingbing.sun@kaust.edu.sa} \\
  \And
  Tariq Alkhalifah \\
  KAUST \\
  Saudi Arabia, Thuwal \\
  \texttt{tariq.alkhalifah@kaust.edu.sa} \\
}

\pagebreak

\date{\vspace{-5ex}}
\begin{document}

\maketitle

\input{Texts/Abstract}
\input{Texts/Introduction}
\input{Texts/TLchallenges}
\input{Texts/LSTM}
\input{Texts/Method}

\input{Texts/Repeatability}
\input{Texts/Examples}
\input{Texts/Noise}
\input{Texts/Discussion}
\input{Texts/Conclusion}

\input{Texts/Appendix}

\bibliographystyle{seg}

\bibliography{ref}
\end{document}

%% file: Texts/Abstract.tex
\begin{abstract}
Time-lapse seismic data acquisition is an essential tool to monitor changes in a reservoir due to fluid injection, such as CO$_2$ injection. By acquiring multiple seismic surveys in the exact location, we can identify the reservoir changes by analyzing the difference in the data. However, such analysis can be skewed by the near-surface seasonal velocity variations, inaccuracy in the acquisition parameters, and other inevitable noise. The common practice (cross-equalization) to address this problem uses the part of the data where changes are not expected to design a matching filter and then apply it to the whole data, including the reservoir area. Like cross-equalization, we train a recurrent neural network on parts of the data excluding the reservoir area and then infer the reservoir-related data. 
The recurrent neural network can learn the time dependency of the data, unlike the matching filter that processes the data based on the local information obtained in the filter window. We demonstrate the method of matching the data in various examples and compare it with the conventional matching filter. Specifically, we start by demonstrating the ability of the approach in matching two traces and then test the method on a pre-stack 2D synthetic data. Then, we verify the enhancements of the 4D signal by providing RTM images. We measure the repeatability using normalized root mean square and predictability metrics and show that in some cases, it overcomes the matching filter approach.
\end{abstract}

%% file: Texts/Introduction.tex
\section*{Introduction}
% general intro 
4D seismic technology, also known as time-lapse (TL) seismic, is the process of repeating the seismic survey many times in the same location with the same acquisition parameters. The period between the surveys ranges from a few months to years, and the data are often acquired before and after operations that might induce changes in the reservoir (e.g., injection). It is essential for monitoring purposes, such as monitoring fluid substitutions and saturation changes in a reservoir and monitoring CO$_2$ injection in carbon sequestration projects (e.g. Sleipner \citep{chadwick2004geological}, Weyburn \citep{preston2005iea}, and  In Salah \citep{ringrose2013salah}). In addition, it can provide indications of the changes in the reservoir thickness resulting from compaction \citep{hatchell2005measuring}. The quality of the TL data depends on the repeatability of the signal and noise level. Ideally, the TL seismic surveys should provide identical TL data except at the target area where the fluid movement occurred. Unfortunately, TL data always have non-repeatable signals due to many factors such as ambient noise, velocity variation in the near-surface, source-receiver positioning, and other survey factors \citep{bakulin2012feasibility,shulakova2014burying,nguyen2015reviewTimeLapse}.
\\
\\
% TL in liturature 
Many researchers have addressed these issues and proposed various techniques to improve the repeatability and reduce the 4D noise. Generally, enhancing the repeatability requires improving the acquisition (e.g., system node technology, permanent acquisition \citep{bakulin2018permanent}, simultaneous 4D pre-stack processing \citep[e.g.,][]{nguyen2015reviewTimeLapse} and inversion \citep[e.g.,][]{kazei2017centered}). Traditionally, 4D processing is implemented by matching the monitor to the baseline data through cross-equalization techniques such as matched-filtering \citep{rickett2001cross,robinson2000geophysical}. To avoid deforming the real differences embedded in the target signals, the matched-filtering operator is computed in a shallow time window so that it does not contain any reservoir signal \citep{rickett2001cross}. The length of the window used to compute the matching filter is critical. A short window can be dominated by harmful noise, while choosing a large window might affect the spatial resolution \citep{lumley20034d,williamson2007new}. 
\\
\\
% DL and TL 
In the age of digital transformation, deep learning models have been used widely in many seismic applications such as data processing \citep{ovcharenko2019deep,kazei2019realistically}, modeling \citep{song2020solving}, inversion \citep{araya2018deep,kazei2020deep,sun2020mlmisfit} and interpretation \citep{xiong2018seismic,sen2020saltnet}. The type of neural network architecture needed depends on the application. For example, convolutional neural network (CNN) architectures are commonly used in classification, segmentation, and image processing tasks, but they are less reliable in learning the time dependency for time series problems. Recurrent neural networks (RNN), on the other hand, are robust in dealing with time dependency tasks and are often used in time series modeling and prediction. Seismic data are time series; therefore, they are suitable for RNN. Some applications of RNN in geophysics include seismic deconvolution \citep{pereg2020sparse}, NMO velocity prediction \citep{biswas2018stacking}, and inversion \citep{richardson2018seismic,adler2019deep,alfarraj2019semisupervised}.
\\
\\
In the context of time-lapse, \cite{yuan2020time} utilized a convolution neural network (CNN) to predict the velocity changes in different vintages. They tested the parallel and double-difference strategies originally developed in traveltime tomography and full-waveform inversion \citep{kamei2017full}. In the parallel implementation, CNN is used to predict the velocity models for the base and the monitor separately. For the double-difference strategy, the network is used directly to predict the model difference from the data difference between the base and the monitor, but it assumed that the data shared the same acquisition and overburden. \cite{zhou2019data} proposed a spatial-temporal dense CNN to map the seismic data to CO$_2$ leakage mass. As the leakage is accumulated with time, they further combined the network with RNN to utilize the time-dependency for the leakage. \cite{timelapse_auto} proposed to use fully connected layers in a latent space of an autoencoder to match the monitor and the base data. \cite{duan2020estimation} trained a network to estimate the time-shift between the base and the monitor in a zero-offset seismic data in a supervised fashion. These studies require generating training datasets that is diverse enough to generalize the network, which is a challenging task as many variables need to be considered.  \cite{jun2021repeatability} improved the repeatability of post-stack seismic images by training the network to remove the non-repeatable noise in the part of the data that do not have reservoir information (like early arrivals), then inferring on the reservoir area.
\\
\\
In this paper, we use long short-term memory (LSTM) network, a type of RNN, to map traces from the base to the monitor. We train the network on shallow window that does not contain the target signal, similar to what has been done using the matching filter approach and then infer for the deeper parts. This training methodology does not require generating training datasets as in \cite{yuan2020time,zhou2019data,duan2020estimation}. It also deals only with seismic data, unlike the work of \cite{yuan2020time} and \cite{zhou2019data} that inherently require the network to learn the physics of mapping the data to velocities and the CO$_2$ mass. Our approach is close to the work implemented by \cite{jun2021repeatability} except they only focus on reducing the 4D noise in a local image batch and their method is prone to fail when a large time shift exists between the vintages.
\\
\\
The paper is organized as follows. We start by providing an overview of the time-lapse challenges. Then, we explain the method to match the data using LSTM. Next, we demonstrate the training strategy and show how we utilize it to correct the time-shift in TL data. After that, we verify the method on synthetic examples. The first example is a simple matching of two traces. The second and third examples are synthetic 2D models based on the Otway and time-lapse SEAM models. Finally, we discuss the results and the method.

%% file: Texts/TLchallenges.tex
\section*{Time-lapse challenges}
The recorded seismic data $d$ can be simplified using the Earth convolutional model, which is given by the convolution of the Earth's response (i.e. reflectivity) $R$ with the source signature $S$ in addition to some ambient noise $n(t)$ and it is written as, 
\begin{equation}
    d(t)  =  S(t) * R(t) + n(t),
    \label{eq:conv}
\end{equation}
where $*$ denotes the convolutional operator (over time), and $t$ is time.  The seismic source $S(t)$ can include the interaction between the seismic wavelet and the near-surface (the effective source). $S(t)$ is highly affected by the changes in the weathering layer such as changes from the seasonal variations including rain and wind. It is also sensitive to source-receiver coupling, positioning and equipment malfunction. The reflectivity $R$ depends on the elastic properties of the Earth including multiples. The noise $n(t)$ can be generated from various factors such as human and environmental activities.
\\
\\
In TL seismic, the signal-to-noise ratio and the repeatability between the base $d_B$ and the monitor $d_M$ are key factors to interpret the 4D signal accurately. However, both $S(t)$ and $n(t)$ are prone to changes resulting in non-repeatable signals. For example, in Saudi Arabia, a TL project, conducted in a desert environment with continuous movements of sand dunes overlying karsted limestone layers, strong surface waves, and high lateral velocity
heterogeneity, result in non-repeatable poor quality data \citep{jervis2018making}. In the Otway project, permanent receivers were deployed to minimize the changes in $S(t)$, but they can be unstable overtime \citep{popik20204d}. 
\\
\\
In fact, all TL projects aim to minimize the non-repeatability. This implies reproducing the same $S(t)$ and eliminating the effect of $n(t)$ such that the only difference between $d_B$ and $d_M$ is the difference due to changing $R(t)$ at the reservoir. Increasing the repeatability includes improving the acquisition technology and the processing workflow. For the acquisition, permanent ocean bottom nodes were deployed in Valhall field \citep{roste2007monitoring} and buried receivers are used in the Otway field and in Saudi Arabia \citep{popik20204d,jervis2018making}. Despite the advancement in acquisition, near-surface seasonal variations and various other factors still affect the repeatability of the signal \citep{TL2017processing}. Simultaneous and joint processing are used for 4D seismics to further improve the signal \citep{roach2015TLassessment}. The former is defined by processing different surveys with an identical workflow, while the latter is when we merge the data to estimate certain processing parameters. During processing, a cross-equalization step is commonly applied in which different data are aligned in time and amplitude.   

%% file: Texts/LSTM.tex
\section*{Theory \& Method}
In TL seismics, we have a base data $d_B$, which is often considered our reference. A fluid, often CO$_2$, is then injected into a reservoir and the monitor data $d_M$ is acquired to observe the changes $R$ due to the injection. The difference $\delta d$ between $d_B$ and $d_M$ can be written as, 
\begin{equation}
    \delta d(t) \, \,  =  d_B(t) - d_M(t) \, \, = N(t) + \delta R(t),
    \label{eq:base}
\end{equation}
where $N(t)$ is the 4D noise resulting from the non-repeatable signal (i.e. changes in $S(t)$ and $n(t)$ (Equation~\ref{eq:conv})), $t$ is the time. The target of the TL experiment is to identify $\delta R$ but this cannot be achieved without eliminating $N$ first. At early time $\delta R(t)=0$, and equation~\ref{eq:base} is simplified to, 
\begin{equation}
    \delta d(t) \, \, =  d_B(t) - d_M(t) \, \,  = N(t).
    \label{eq:2}
\end{equation}
Training a network in this range to map $d_B$ to $d_M$ will teach the network to compensate for the noise $N$. After that, we can use the trained network on the full record and obtain the reservoir changes $\delta R$.   

\subsection*{Long short-term memory (LSTM)}
\begin{figure}
    \centering
    \includegraphics[width=0.99\textwidth]{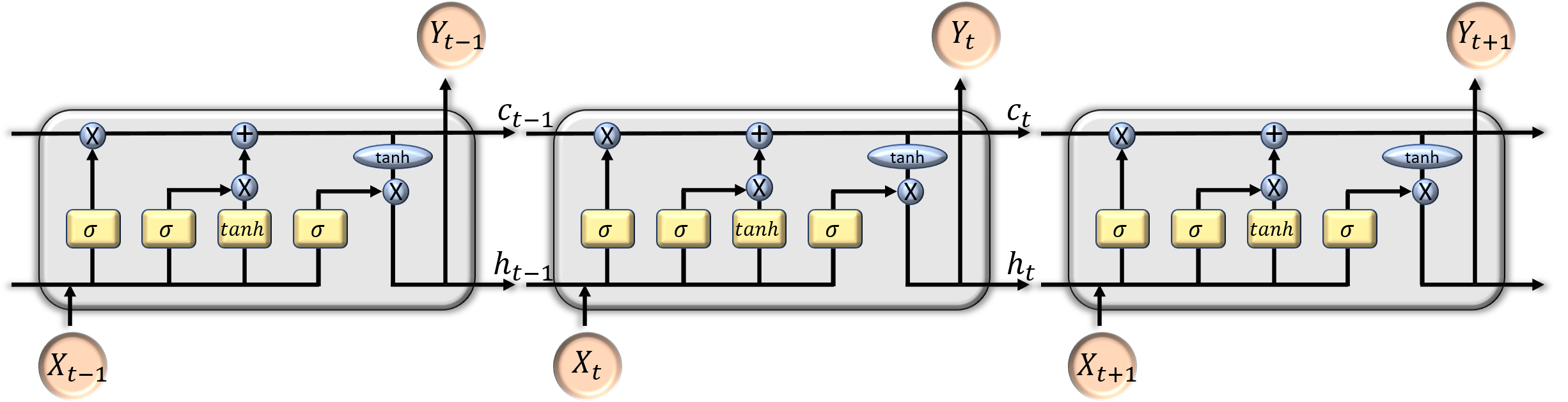}
    \caption{LSTM unit, $h_t$ and $c_t$ are the hidden and cell state, respectivvely. $X_t$ and $Y_t$ represent the input and the output at time sample $t$.}
    \label{fig:LSTM}
\end{figure}

RNN networks are robust for time series data as they contain a feedback loop that carries information from previous time steps to the current RNN unit. For long sequences, the feedback loop can depend on many past time-steps resulting in gradient decay, commonly known as the vanishing gradient problem. LSTM, a variant of RNN, includes a hidden state ($h_t$), a cell state ($c_t$) and gating mechanisms to maintain the long-term dependency without falling into the vanishing gradient issue. More details about the structure of LSTM is given in appendix A.
% An LSTM unit consists of three gates: A Forget gate ($f_t$), that removes the useless information, an Input gate ($i_t$), which updates the internal states, and an Output gate ($o_t$), that forms the final output. The gates are composed of fully-connected neural networks combined with activation functions. Mathematically, LSTM updates the cell and hidden state as follow:  
% \begin{align}
%     f_t &= \sigma(W_f x_t + U_f h_{t-1} + b_f) \\
%     i_t &= \sigma(W_i x_t + U_i h_{t-1} + b_i)\\
%     o_t &= \sigma(W_o x_t + U_o h_{t-1} + b_o)\\
%     g_t &= tanh (W_g x_t + U_g h_{t-1} + b_g)\\
%     c_t &= f_t \bullet c_{t-1} + i_t \bullet g_t\\
%     h_t &= o_t \bullet tanh(c_t), 
% \end{align}
% Where $W$ and $U$ are matrices for the weights and the recurrent connection, and $b$ is a bias vector. $\sigma$ represents the sigmoid activation function and ($\bullet$) is an element-wise multiplication.
\\
\\
The network is trained sequentially to learn the time dependency: a temporal sequence ($x$) is fed to the network from $t=0$ to $t=N$, where $N$ is the length of the sequence. In Figure~\ref{fig:LSTM}, we show the flow of an LSTM unit. When LSTM receives a temporal input ($x_{t}$), it will update the cell state ($c_{t-1}$) and the hidden state ($h_{t-1}$), which are saved from the previous time samples. The updated internal state ($c_{t}$,$h_{t}$) are used for the next temporal input ($x_{t+1}$) and so on. The output of LSTM can be a single sample (e.g., predicting the opening stock price based on the previous days) or a sequence (e.g., language translation). 

%% file: Texts/Method.tex
\subsection*{Training methodology}
Here, we propose to utilize LSTM to predict the behavior of the monitor from the base data. To train the network, we use traces from the base as the inputs. The input trace is further divided into overlapping windows of size $w$ making the dimension of the input equal to \{$Batch\, size,n_t,w$\}, where $n_t$ is the time samples. The output is the middle value of the window $w$ of the corresponding trace from the monitor with the shape (($Batch\, size,n_t,1$)). We split the trace into two parts: a shallow part that does not contain any reservoir signal and a deeper part containing the reservoir. The training is implemented in the first part only. This allows the network to learn the difference between the base and the monitor resulted from the near-surface variation only, then, correct it. The loss we use is the mean squared error (MSE) written as:
\begin{equation}
    MSE = \frac{1}{N}\sum_{i=1}^N (\theta(b_i)-m_i)^2, 
\end{equation}
where $\theta$ represent the network, b and m are the shallow part of traces from the base and the monitor, respectively, and $N$ is the number of training traces.
After the training, the network is used to infer the traces including the deeper part to obtain the predicted monitor $\bar{M}$,
\begin{equation}
    \bar{M}=\theta(B).
\end{equation}
We use capital letters $B$ and $\bar{M}$ to indicate that all the traces are used. Subtracting the predicted monitor $\bar{M}$ with the actual monitor $M$ should yield zero except at the reservoir because the network did not learn how to map the reservoir changes. 
\\
\\
Some pre-processing to the data is implemented before starting the training. First, we mute the direct arrivals and the diving waves. They do not share the same wave path as the target reflections from 4D changes. Hence, including the direct arrivals and the diving waves would negatively impact the corrections for the target horizon. Second, scaling the data by using standard scaling or Min-Max scaling is necessary for convergence.

%% file: Texts/Repeatability.tex
\section*{Repeatability metrics}
Repeatability metrics are quantitative measurements to assess the quality of the repeatibility obtained in a 4D seismic experiment. In this paper, we will use two repeatability metrics that are the normalized root mean square (NRMS) and the predictability. 
\\
\\
NRMS is sensitive to changes in amplitude and phase in the data \citep{kragh2002nrms}. It is implimented by normalizing the average of the difference RMS by the average RMS energy, and it reads, 
 \begin{equation}
     NRMS = \frac{200  * RMS(B-M)}{RMS(B)+RMS(M)},
 \end{equation}
 where B and M represent a window from the baseline and the monitor data, respectively. Its values ranges from 0\% to 200\% and the smaller the NRMS is the better the repeatability. 
\\
\\
The predictability (PRED) measures the correlation between the reflectors, but it is insensitive to the amplitude and phase \citep{waage2019repeatability}. It gives a value from 0\% to 100\% where 100\% means that the traces are correlated. It is expressed by the summed square of the cross-correlation divided by the summed products of the auto-correlation of the two traces within a time window, which yields,  
\begin{equation}
    PRED = \frac{\sum \phi_{bm}^2(\tau) }{\sum \phi_{bb}(\tau) \cdot \phi_{mm}(\tau)},
\end{equation}
where $\phi_{ab}$ is the normalized cross-correlation given by, 
 \begin{equation}
         \phi_{ab}(\tau) = \frac{\sum_{i=w-\tau}^{w+\tau} a(i) \cdot b(i) }{\sqrt{\sum_{i=w-\tau}^{w+\tau} a(i)^2 \sum_{i=w-\tau}^{w+\tau} b(i)^2}},
 \end{equation}
and $w$ indicates the time window.

%% file: Texts/Examples.tex
\section*{Experiments}
We validate the proposed method using three experiments. The first one is an elementary example, which demonstrates the basic idea for using LSTM for matching two traces of data. The second and third examples are synthetic time-lapse data generated using the acoustic constant-density wave equation. For both examples, we simulate the monitor data by including a reservoir variation and smooth random velocity perturbations with 50 m/s mean and 100 m/s standard deviation in the first 20 m depth of the baseline velocity model to mimic realistic near-surface variations.  

\subsection*{Matching two traces}
\begin{figure}
	\centering
	\subfigure[]{\includegraphics[width=0.8\columnwidth]{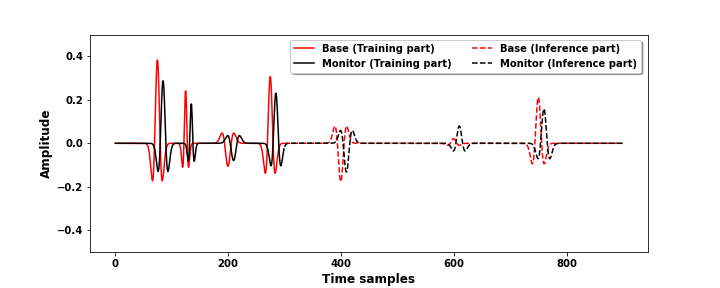}
	\label{fig:2traces_1}}
	\hspace*{-0.01\columnwidth}
	\subfigure[]{\includegraphics[width=0.8\columnwidth]{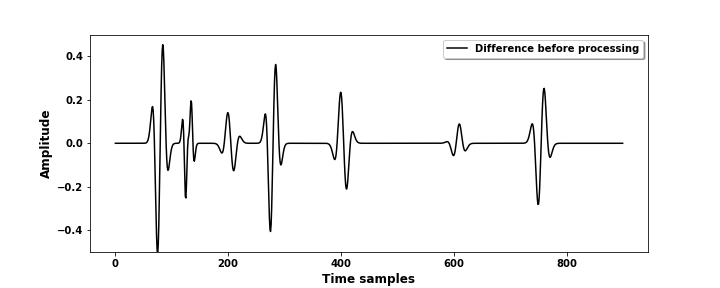}
	\label{fig:2traces_2}}
	\vspace*{-0.01\columnwidth}
    \label{fig:2traces}
    \caption{(a) is a base trace (red line) and a monitor trace (black line). The solid line indicates the training region. (b) shows the difference between the two traces.}
\end{figure}

\begin{figure}
	\centering
	\subfigure[]{\includegraphics[width=0.8\columnwidth]{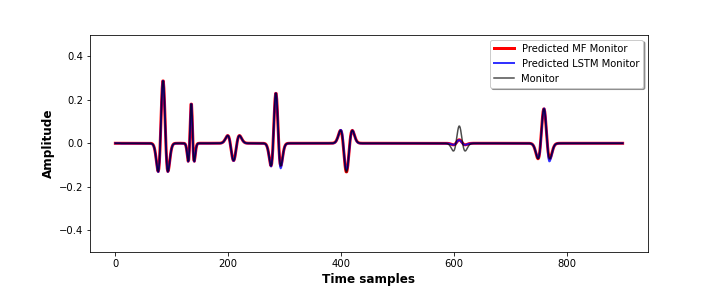}
	\label{fig:2trace_proc_1}}
	\hspace*{-0.01\columnwidth}
	\subfigure[]{\includegraphics[width=0.8\columnwidth]{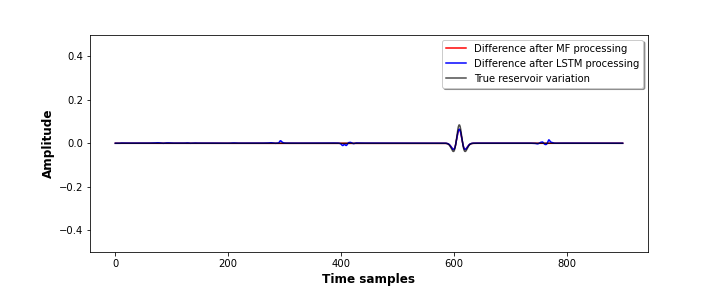}
	\label{fig:2trace_proc_2}}
	\vspace*{-0.01\columnwidth}
    \caption{(a) is the predicted monitor by LSTM ( blue line) and by the matching filter (red line) plotted over the monitor (black line). (b) is the differences between predicted monitors and the actual monitor, as well as, the true reservoir variation. The matching filter almost perfectly matched the monitor as this is a simple example. The LSTM processing successfully matched the two traces with minor artifacts.}
    \label{fig:2trace_proc}
\end{figure}

In this experiment, we show how LSTM is capable of learning a time-shift between two traces. We use a dummy trace with few wiggles as the base. We create a monitor trace by shifting the base and reduce its the amplitude by a scalar. The two traces are shown in Figure~\ref{fig:2traces_1}. We add a small value to the wiggle at sample 600 in the monitor to represent a reservoir variation. We use only the first 300 time samples to train the model to map the base to the monitor by an LSTM layer. Then, we infer for the whole trace. As explained in the method, the input consists of overlapping window of size $w$. We choose $w$ in this example to be 20 samples. We also apply the conventional matching filter with a filter size equal to $w$. We plot the result of applying the method in Figure~\ref{fig:2trace_proc}. Figure~\ref{fig:2trace_proc_1} shows the predicted monitor by LSTM (blue line) and by matching it with the conventional matching filter (red line) with the actual monitor (black line). Figure~\ref{fig:2trace_proc_2} is the reservoir variation after taking the difference between the predicted and the actual monitor. We see that the matching filter approach accurately aligns the two traces as this is a simple example. LSTM successfully aligns the two traces but it has minor artifacts. 
\subsection*{Otway model}
\begin{figure}
	\centering
	\subfigure[]{\includegraphics[width=0.7\columnwidth]{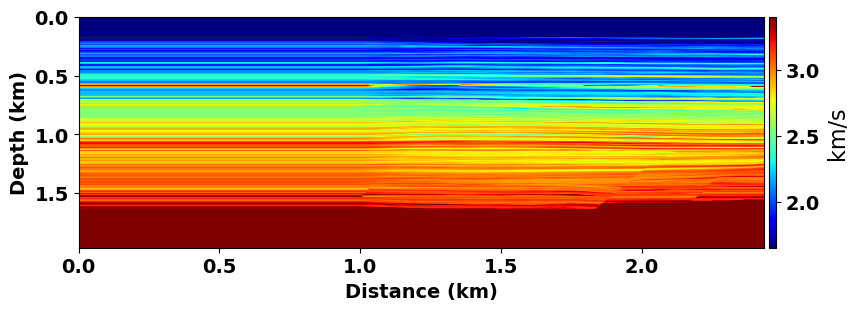}
		\label{fig:Otw_base_model}}
	\vspace*{0.05\columnwidth}
	\subfigure[]{\includegraphics[width=0.7\columnwidth]{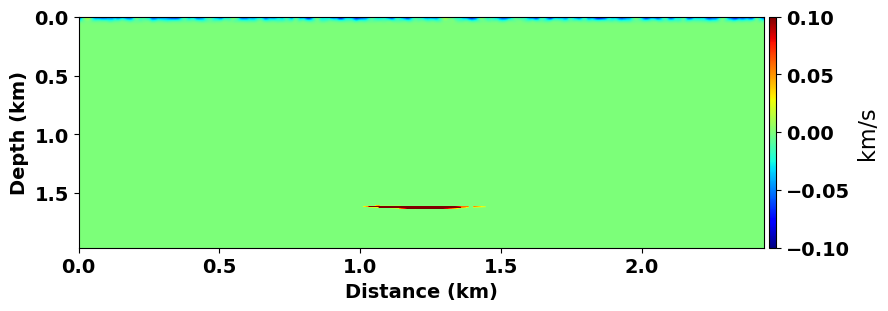}
		\label{fig:Otw_static_model}}
	\vspace*{0.05\columnwidth}
	\subfigure[]{\includegraphics[width=0.7\columnwidth]{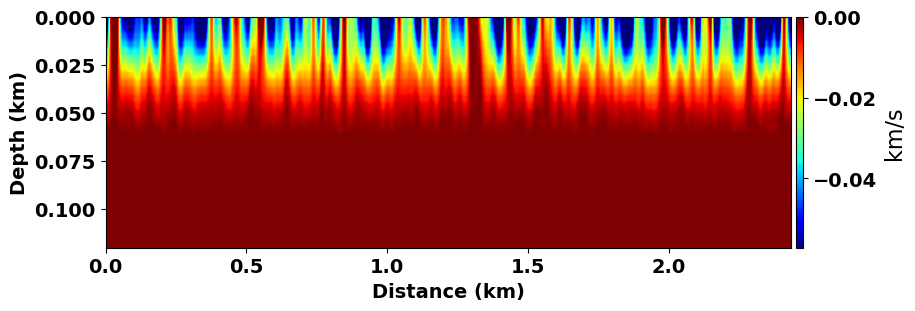}
		\label{fig:Otw_zoom_noize}}
	\caption{(a) The Otway velocity model \citep{glubokovskikh2016OtwayRef} used to generate data for the baseline, (b) the reservoir signal and near-surface variations added to the baseline for the monitor data, and (c)  zoom into the random Gaussian 4D perturbations with 50 m/s mean and 100 m/s standard deviation changes in the first 20 m depth.}
    \label{fig:Otway_vel}
\end{figure}

\begin{figure}
	\centering
	\subfigure[]{\includegraphics[width=0.3\columnwidth]{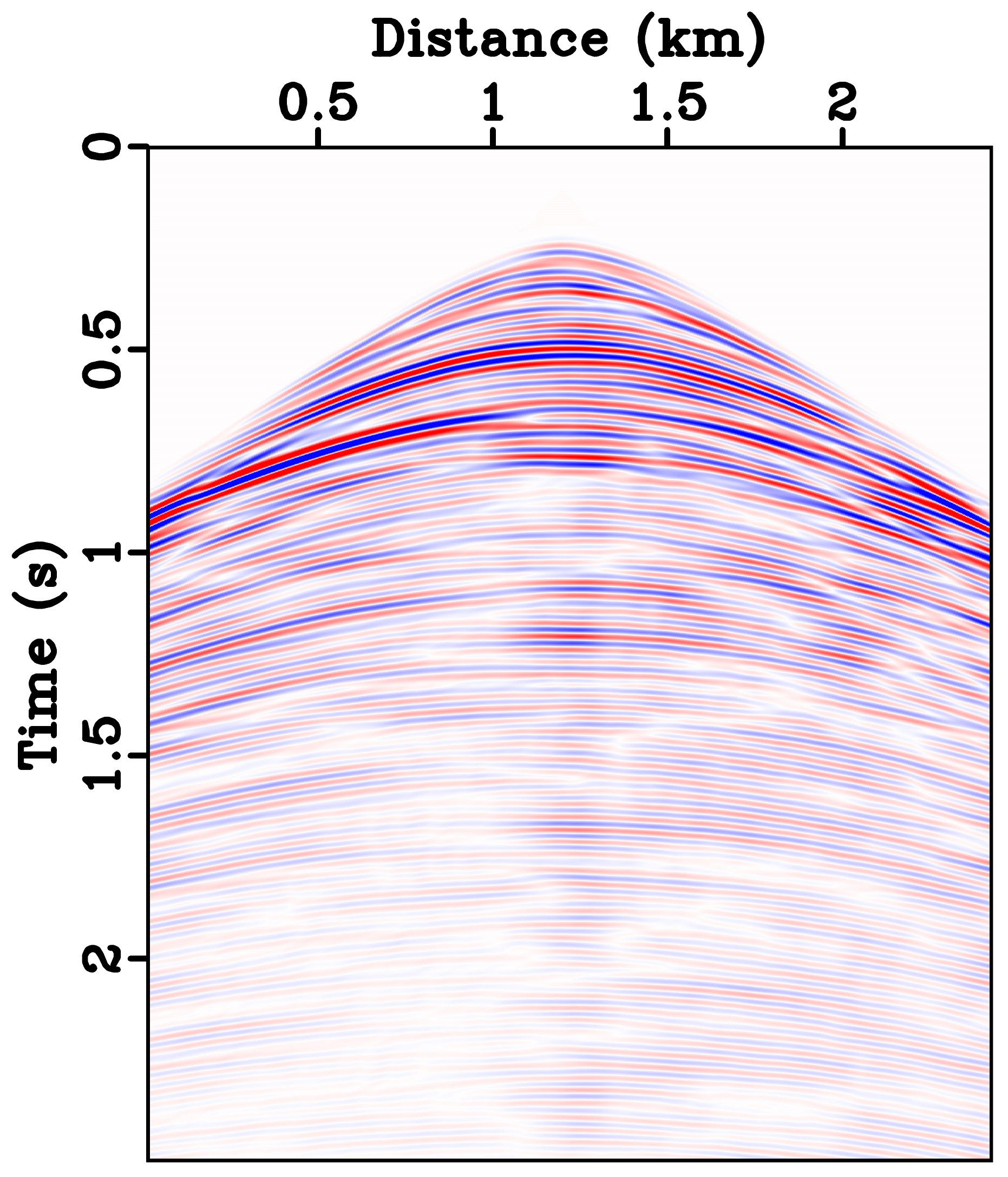}
		\label{fig:Otway_base_shot}}
% 	\vspace*{0.05\columnwidth}
	\subfigure[]{\includegraphics[width=0.3\columnwidth]{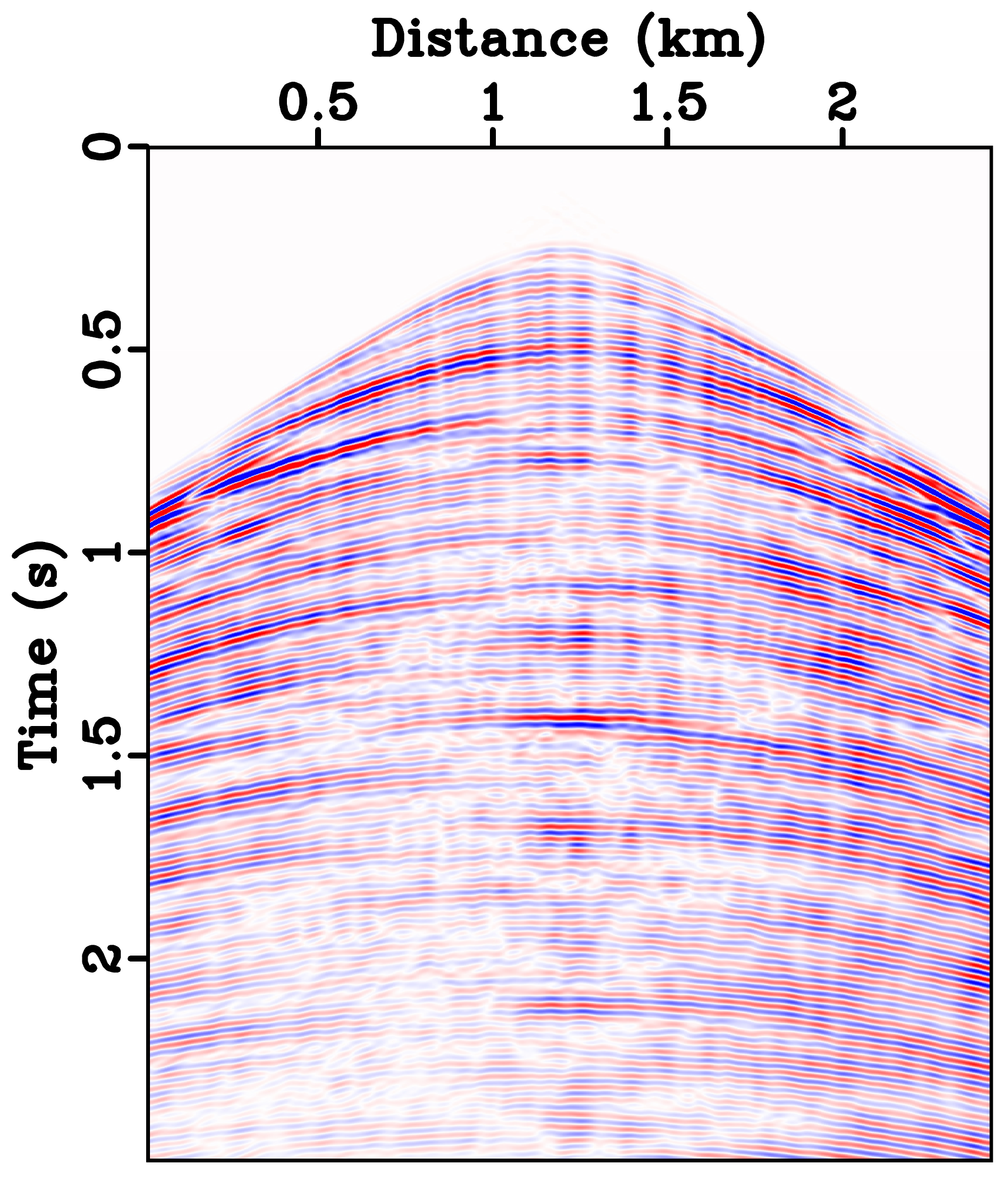}
		\label{fig:Otway_diff_shot}}
% 	\vspace*{0.05\columnwidth}
	\subfigure[]{\includegraphics[width=0.3\columnwidth]{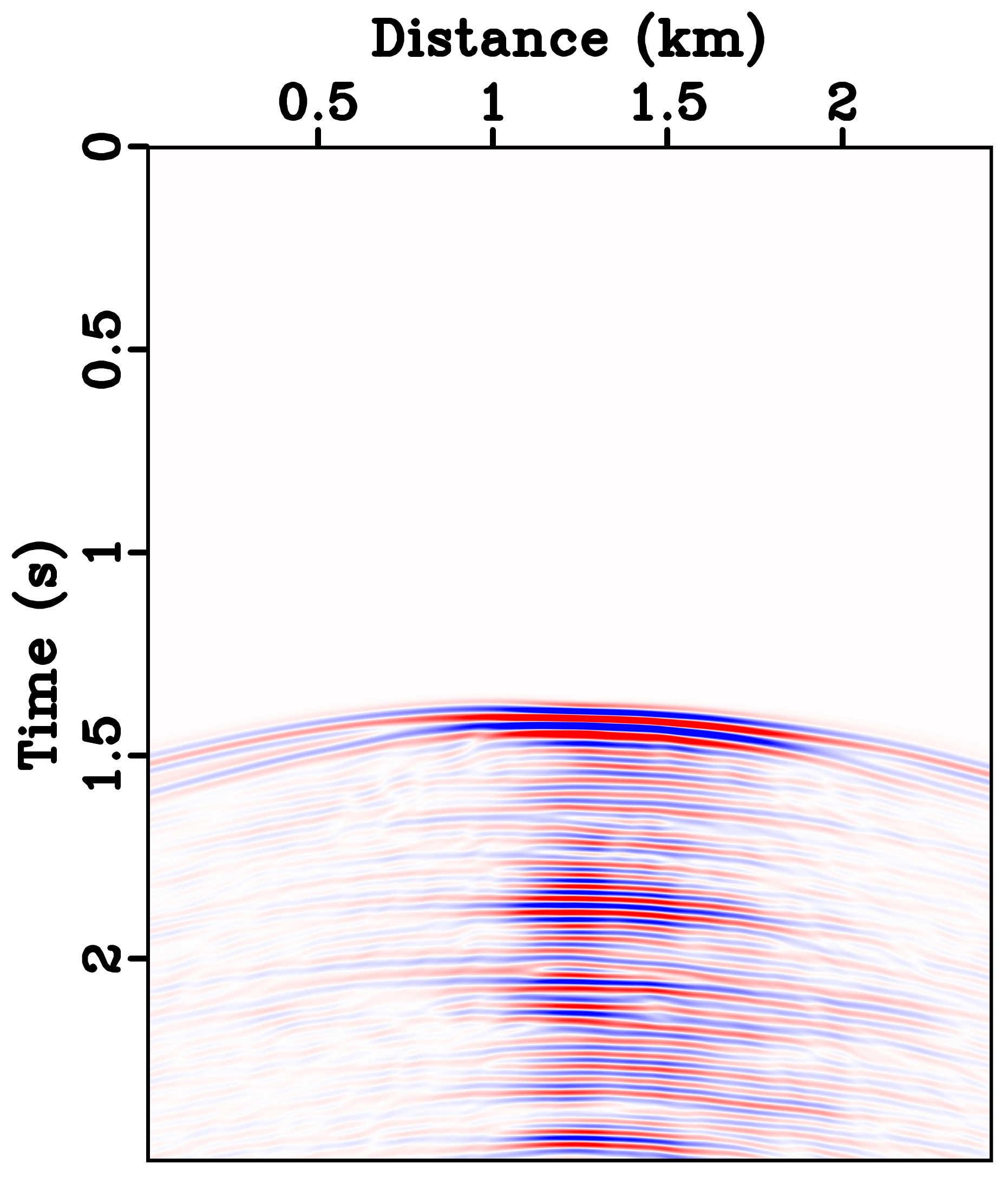}
	    \label{fig:Otway_4d_shot}}
	\caption{A shot gather from the Otway model (a), the data difference between the baseline and the monitor (b) and the target difference (c). The target difference is obtained by taking the difference between the base and a monitor data excluding the near-surface variation in Figure~\ref{fig:Otw_zoom_noize}}
    \label{fig:Otway_shots}
\end{figure}

\begin{figure}
    \centering
    \includegraphics[width=0.6\columnwidth]{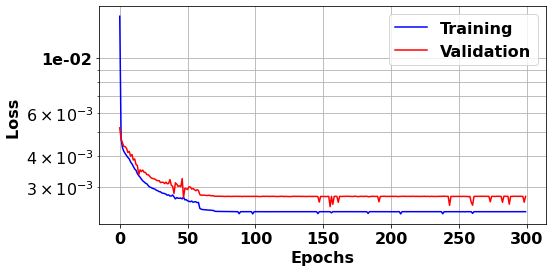}
    \caption{The convergence curve of the MSE loss for Otway data example.}
    \label{fig:loss}
\end{figure}

\begin{figure}
    \centering
    \subfigure[]{\includegraphics[width=.9\textwidth]{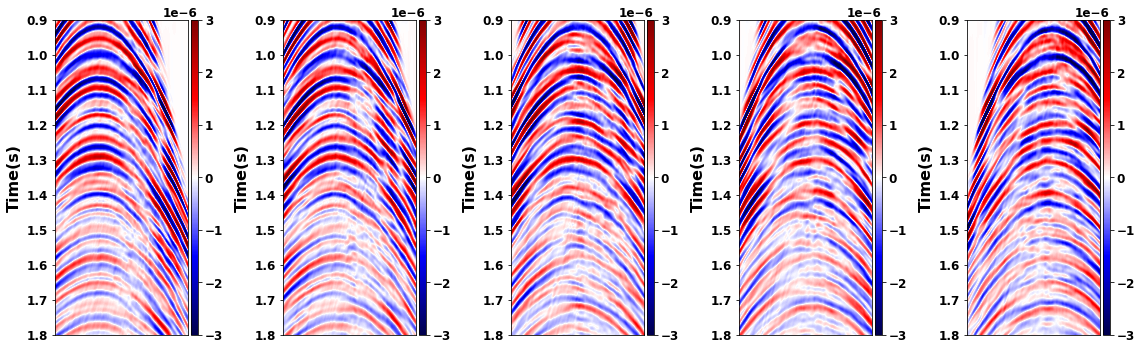} }
    \caption{Shot gathers from the predicted Otway monitor data.}
    \label{fig:pred_mon_otw}
\end{figure}

\begin{figure}
    \centering
    \subfigure[]{\includegraphics[width=.8\textwidth]{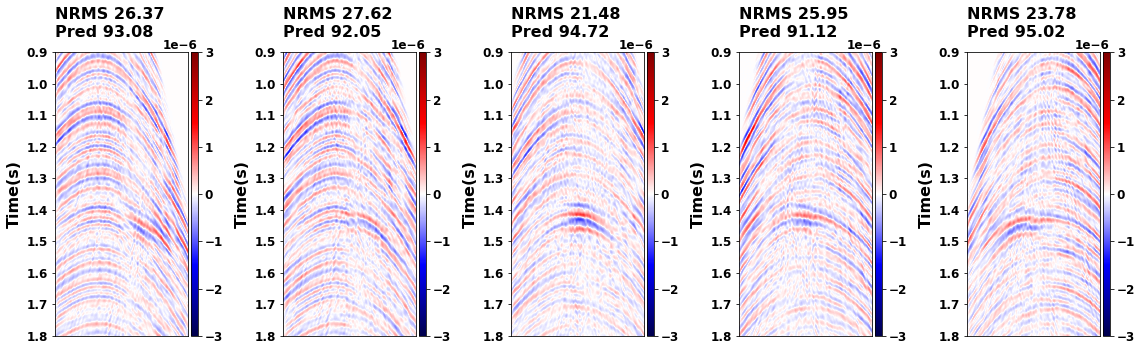} \label{fig:diff_before_otw}}
     \vspace{.1cm}
    \subfigure[]{\includegraphics[width=.8\textwidth]{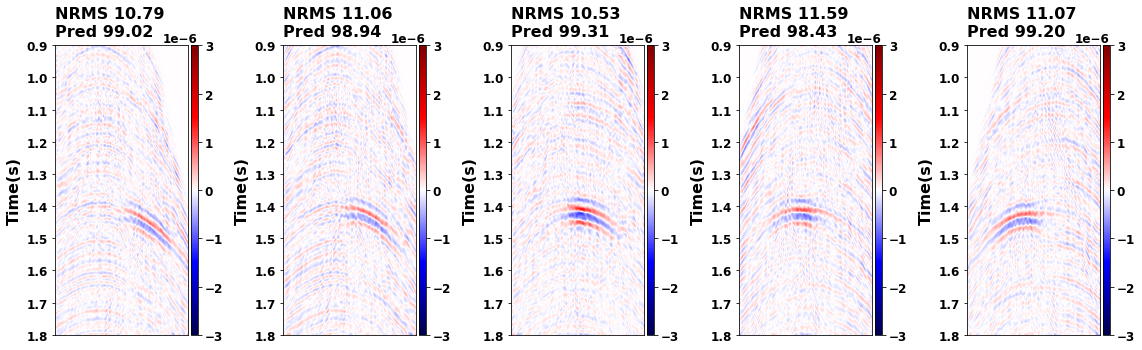} \label{fig:mf_otw}} 
     \vspace{.1cm}
    \subfigure[]{\includegraphics[width=.8\textwidth]{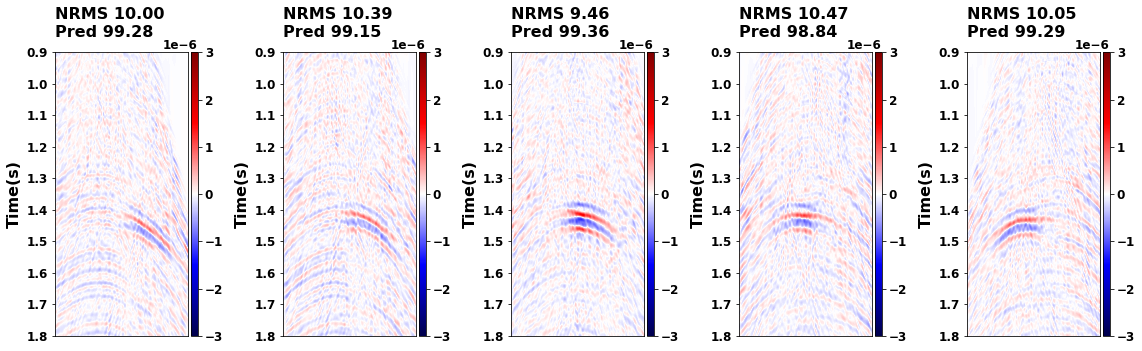} \label{fig:lstm_otw}}
    \subfigure[]{\includegraphics[width=.8\textwidth]{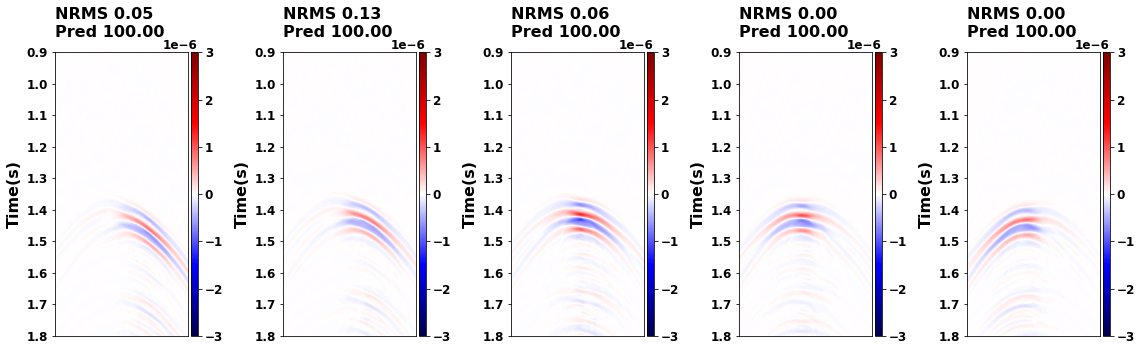} \label{fig:true_diff_otw}}
    \caption{Shot gathers from Otway model for the differences between the monitor and the base. (a) the difference before any processing. (b) is the difference after the matching filter processing. (c) is the difference after the LSTM processing. (d) is the difference without including the near-surface changes corresponding to the true reservoir variations. The NRMS and the predictability are measured in the range 0.9-1.3 s and are displayed at the top of each shots. LSTM processing shows better NRMS and predictability than the matching filter processing in all the shots.}
    \label{fig:shots_diff}
\end{figure}

\begin{figure}
    \centering
    \subfigure[]{\includegraphics[width=.7\textwidth]{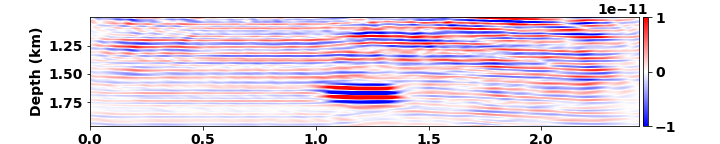}
     \label{fig:before_img_otw}}\\
     \vspace{.1cm}
    \subfigure[]{\includegraphics[width=.7\textwidth]{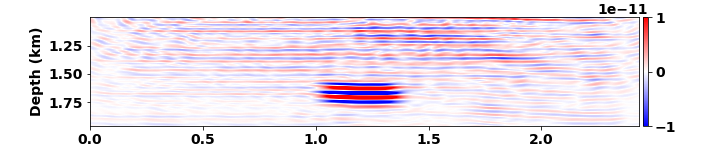}
    \label{fig:mf_img_otw}}\\
     \vspace{.1cm}
    \subfigure[]{\includegraphics[width=.7\textwidth]{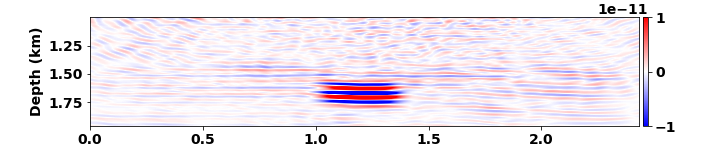}
    \label{fig:lstm_img_otw}}
    \subfigure[]{\includegraphics[width=.7\textwidth]{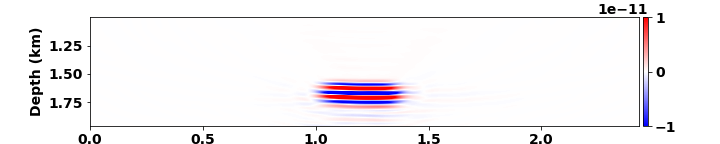}
    \label{fig:true_img_otw}}
    \caption{Otway RTM images of the reservoir variation using the data differences between (a) the monitor and the baseline without processing, (b) after processing with the matching filter approach, (c) after processing with LSTM and (d) the true reservoir  variation}.
    \label{fig:img_otw}
\end{figure}

The Otway velocity model, given in Figure~\ref{fig:Otw_base_model}, is based on a field in Australia and consists of horizontal layers with faults. We show the 4D reservoir signal and the random near-surface variations used to synthesize the monitor data in Figures~\ref{fig:Otw_static_model} and \ref{fig:Otw_zoom_noize}, respectively. We ignited 121 shots separated by 20 m with a 25 Hz Ricker wavelet. The recording length is 2.5 s with a 2 ms sampling rate and a maximum offset of 1.2 km. We display a shot gather from the base in Figure~\ref{fig:Otway_base_shot}. We also show the difference between the base and the monitor in Figure~\ref{fig:Otway_diff_shot} and the target reservoir variation in Figure~\ref{fig:Otway_4d_shot}. The target reservoir variation is obtained by taking the difference between the base and monitor data that does not include the near-surface variation (Figure~\ref{fig:Otw_zoom_noize}). Although, the difference introduced between the base and the monitor is very subtle, it generates enough 4D effect to damage the reservoir signal.
\\
\\
We tested different hyper-parameters and chose the ones that provide the least mean squared error in the validation set. We use two LSTM layers with size 50 neurons, followed by a linear layer with tanh activation to project the LSTM output to the dimension of the target sequence. The reservoir perturbation appears at about 1.4 s as shown in Figure~\ref{fig:Otway_4d_shot}. We create the input data by splitting each input trace with $w=40$ time-sample window from 0.9 s to 1.3 s (200 time sample) as the reservoir response has not yet been recorded at this range. The total number of traces is 24360 divided into 80\% training and \%20 validation sets. The batch size for each iteration is 64 traces. We optimize the network using Adam with a learning rate of 0.002 and we use the mean square error (MSE) loss. We run 300 epochs and plot the convergence curve in Figure~\ref{fig:loss}. Both the training and the validation sets converge almost to the same error.
\\
\\
After training the data on the shallow part of the traces, we infer on the whole data, including the reservoiar region. We show five predicted shot gathers by the network from 1.2 to 1.8 in Figure~\ref{fig:pred_mon_otw}. Figure~\ref{fig:shots_diff} shows the differences of the corresponding shots between the monitor and the base without processing (Figure~\ref{fig:diff_before_otw}), after processing with the matching filter (Figure\ref{fig:mf_otw}) and the LSTM approach (Figure~\ref{fig:lstm_otw}), and the difference without the near-surface variation showing the target reservoir variation in Figure~\ref{fig:true_diff_otw}, all plotted at the same scale. Note that the size of the filter in the matching filter approach is equivalent to the size of $w$ in the LSTM input. Above the shots we display the NRMS and the predictability metrics computed in the range 0.9-1.3 s. By examining the differences, we can clearly see that the processed difference cleans most of the unwanted events that exist between the base and the unprocessed monitor. Comparing Figures~\ref{fig:mf_otw} and \ref{fig:lstm_otw}, we see that the LSTM achieved a slightly better repeatability than the matching filter. 
\\
\\
To further assess the enhancement in the 4D signal, we use the data differences shown in Figure~\ref{fig:shots_diff} to implement reverse time migration (RTM) and image the reservoir variation. RTM images are provided in Figure~\ref{fig:img_otw} for (a) the image before the processing, (b) after applying the matching filter, (c) after applying the proposed approach and (d) the true 4D signal. We can see that the matching filter and the LSTM methods both managed to reduce the unwanted differences resulting from the added near-surface variation. However, the image after the LSTM approach (Figure~\ref{fig:lstm_img_otw}) shows less artifacts than that of the matching filter in Figure~\ref{fig:mf_img_otw}. To validate this claim, we use the structural similarity index measure (SSIM) \citep{wang2004image_ssim} relative to the true variation image, which gives a value between -1 and 1 with 1 indicating that the two images are identical. We found that SSIM is 0.12 before processing, 0.42 after processing with the matching filter and 0.50 after processing with LSTM. This proves that the image after applying LSTM is much similar to the true variation image than the image processed with the matching filter. 
\\
\\
%
% \begin{figure}[!htb]
%     \centering
%     \includegraphics[width=.7\columnwidth]{Fig/Otway/smooth.eps}
%     \caption{A smoothed version of Otway model used as a background velocity for RTM.}
%     \label{fig:smooth}
% \end{figure}
%
\subsection*{SEAM time-lapse model}
\begin{figure}
	\centering
	\subfigure[]{\includegraphics[width=0.7\columnwidth]{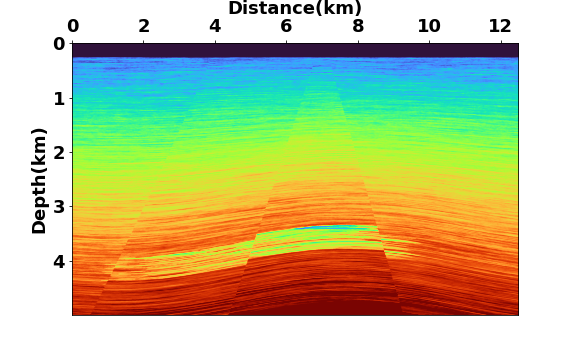}
		\label{fig:seam_base_model}}
	\vspace*{0.05\columnwidth}
	\subfigure[]{\includegraphics[width=0.7\columnwidth]{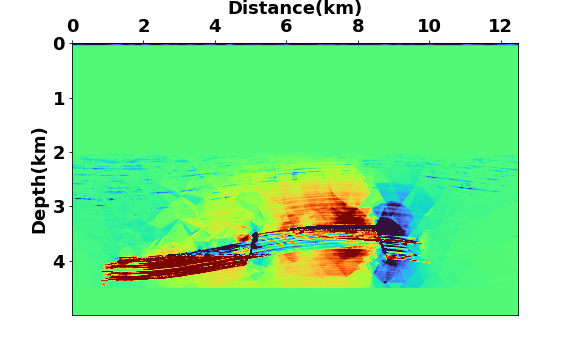}
		\label{fig:seam_diff_model}}
	\caption{(a) The SEAM time-lapse velocity model \citep{oristaglio2016seam} used to generate data for the baseline, (b) the reservoir variations.}
    \label{fig:Seam_vel}
\end{figure}

\begin{figure}
	\centering
	\subfigure[]{\includegraphics[width=0.3\columnwidth]{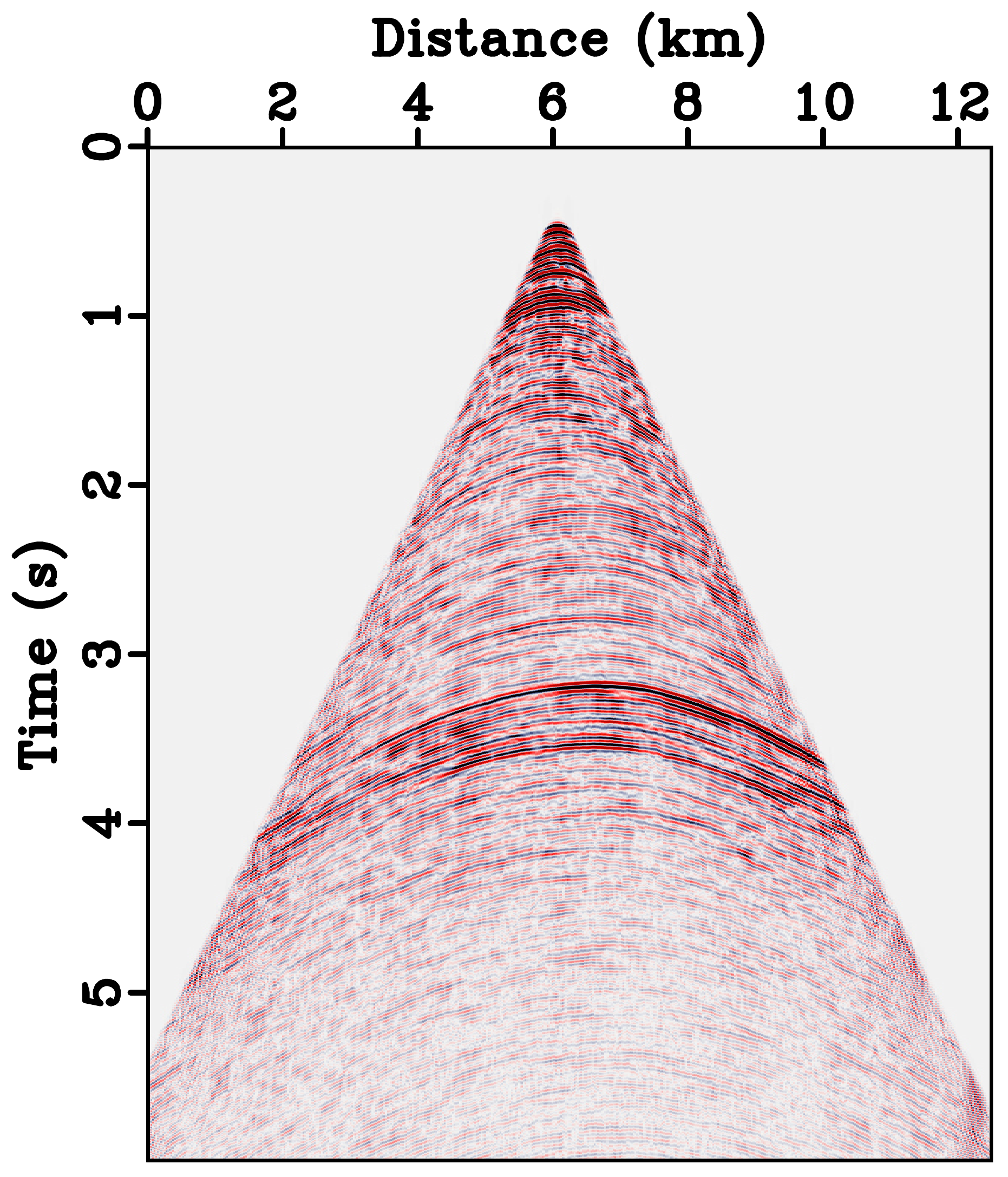}
		\label{fig:seam_base_shot}}
% 	\vspace*{0.05\columnwidth}
	\subfigure[]{\includegraphics[width=0.3\columnwidth]{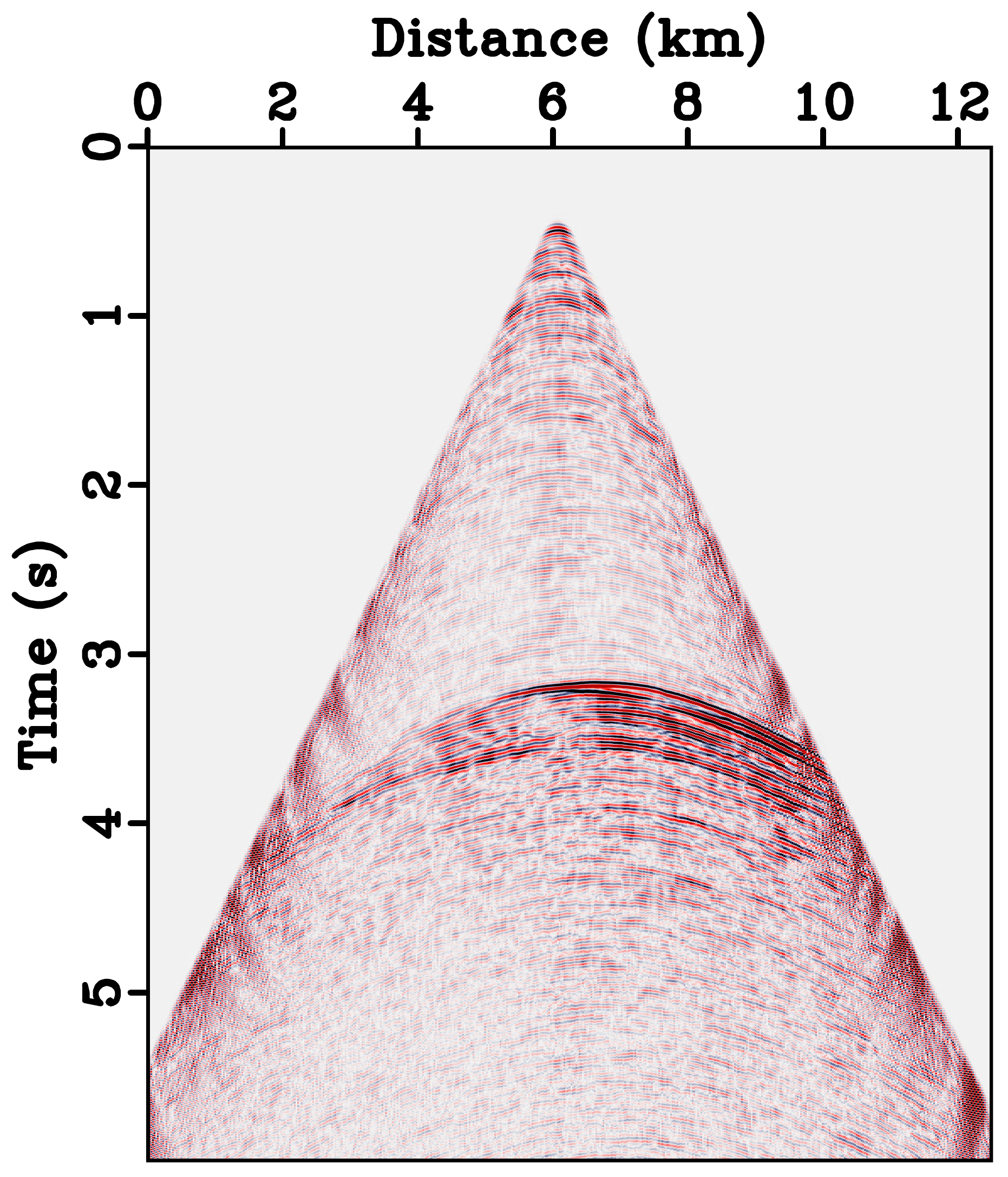}
		\label{fig:seam_diff_shot}}
% 	\vspace*{0.05\columnwidth}
	\subfigure[]{\includegraphics[width=0.3\columnwidth]{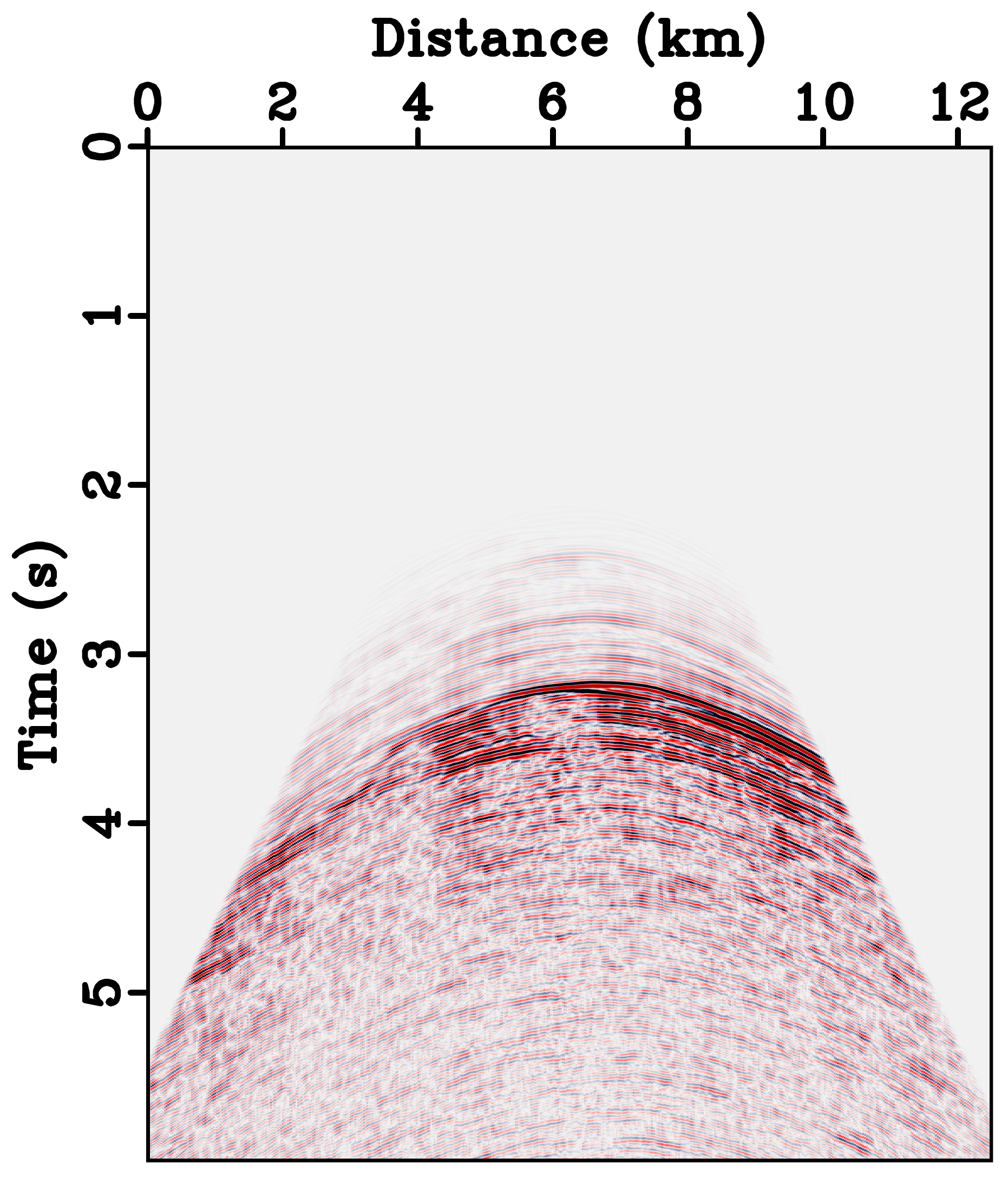}
	    \label{fig:seam_4d_shot}}
	\caption{a shot gather from the SEAM model (a), the data difference between the baseline and the monitor (b) and the target difference (c).}
    \label{fig:Seam_shots}
\end{figure}

\begin{figure}
    \centering
    \includegraphics[width=0.6\columnwidth]{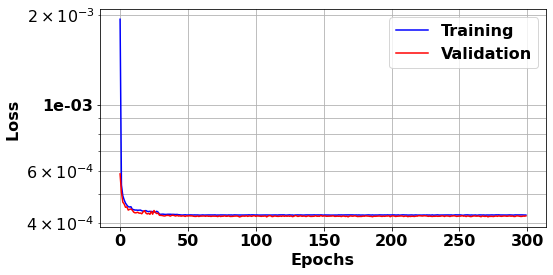}
    \caption{The convergence curve of the MSE loss for SEAM data example.}
    \label{fig:loss2}
\end{figure}

\begin{figure}
    \centering
    \subfigure[]{\includegraphics[width=.9\textwidth]{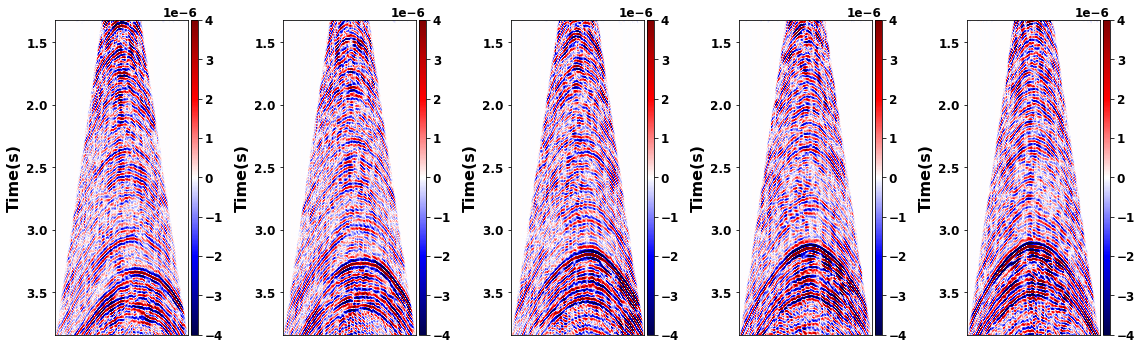}}
    \caption{Shot gathers from the predicted SEAM monitor data.}
    \label{fig:pred_mon_seam}
\end{figure}

\begin{figure}
    \centering
    \subfigure[]{\includegraphics[width=.8\textwidth]{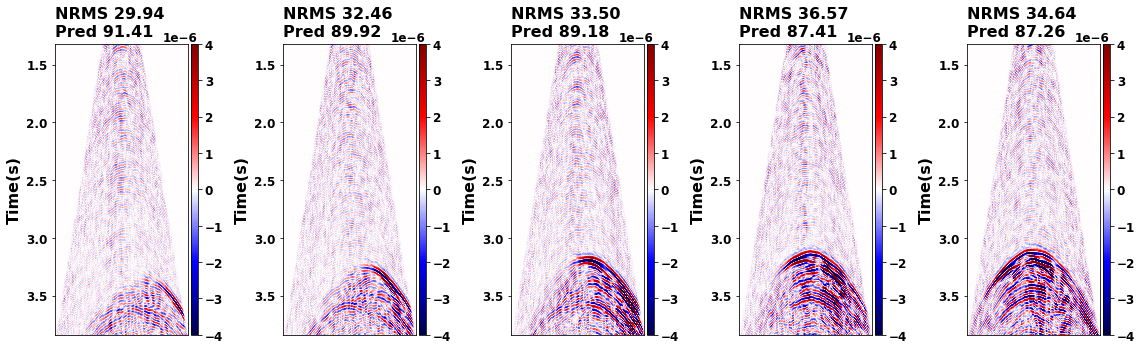} \label{fig:diff_before_seam}}
     \vspace{.1cm}
    \subfigure[]{\includegraphics[width=.8\textwidth]{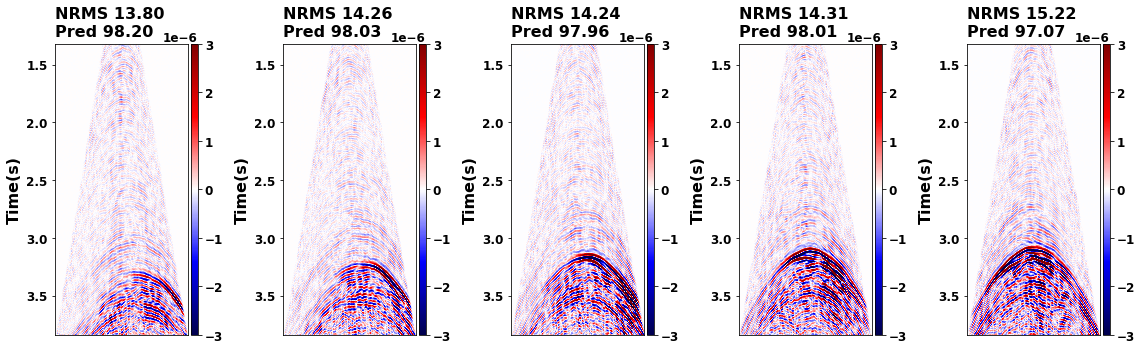} \label{fig:mf_seam}} 
     \vspace{.1cm}
    \subfigure[]{\includegraphics[width=.8\textwidth]{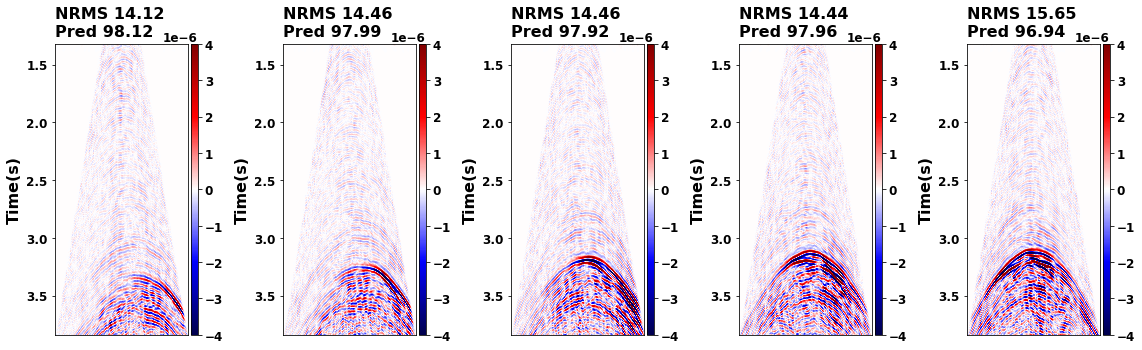} \label{fig:lstm_seam}}
    \subfigure[]{\includegraphics[width=.8\textwidth]{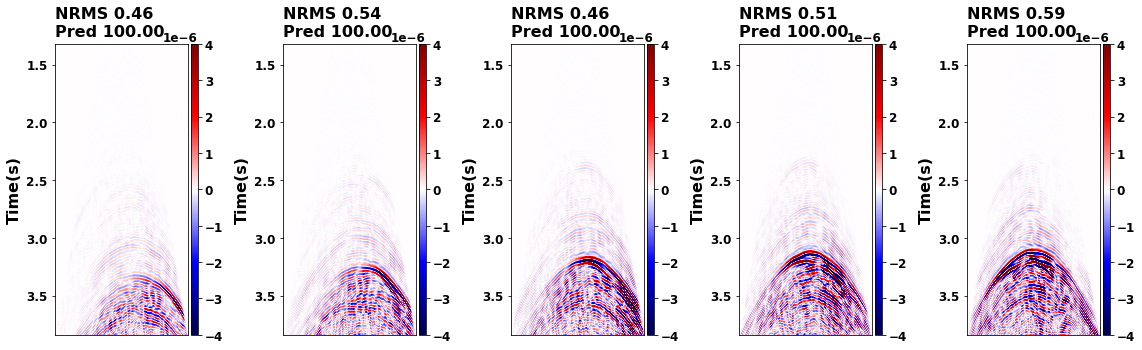} \label{fig:true_diff_seam}}
    \caption{Shot gathers from the SEAM model for the differences between the monitor and the base. (a) the difference before any processing. (b) is the difference after the matching filter processing. (c) is the difference after the LSTM processing. (d) is the difference without including the near-surface changes corresponding to the true reservoir variations. The NRMS and the predictability are measured in the range 1.3 to 2.2 s and are displayed at the top of each shots. The NRMS and the predictability for the LSTM method and the matching filter are similar.}
    \label{fig:shots_diff_seam}
\end{figure}

\begin{figure}
    \centering
    \subfigure[]{\includegraphics[width=.6\textwidth]{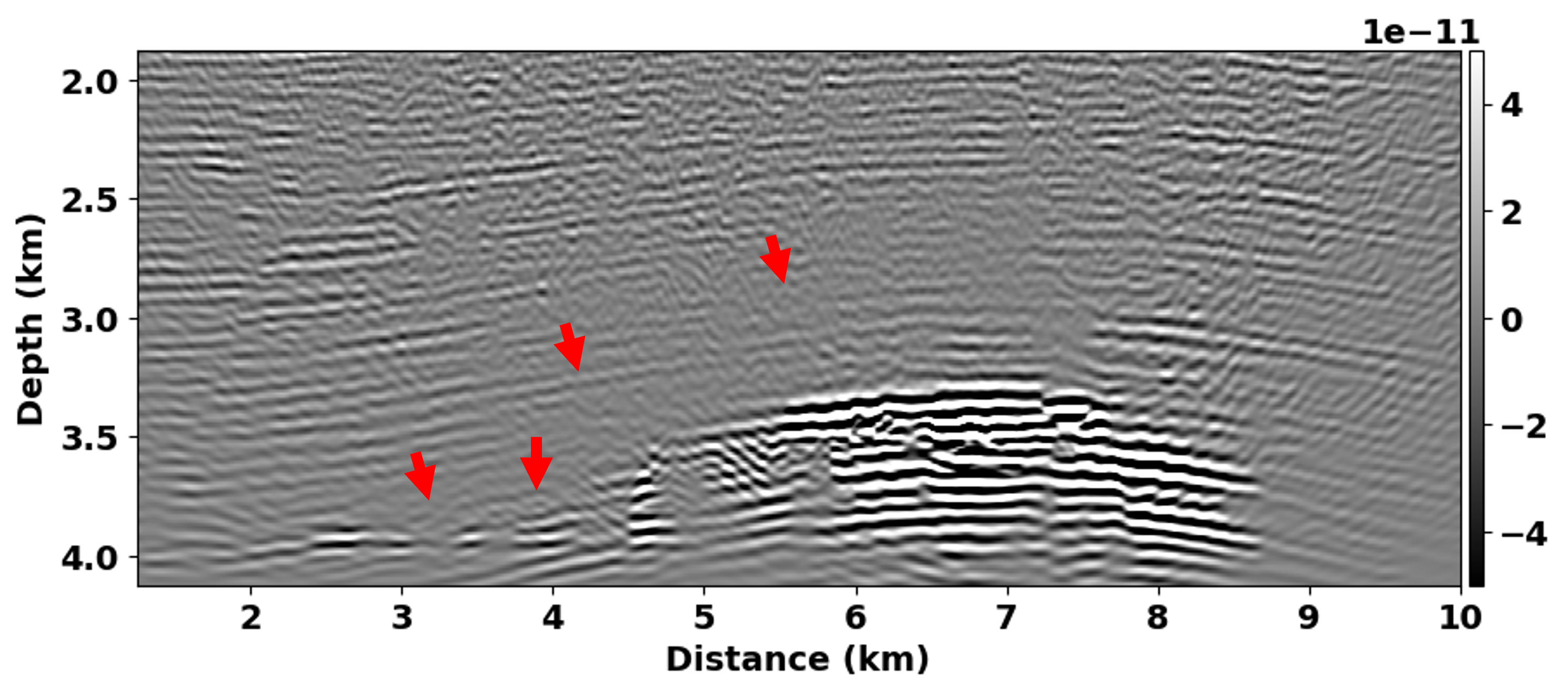}
     \label{fig:before_img_seam}}
     \vspace{.1cm}
    \subfigure[]{\includegraphics[width=.6\textwidth]{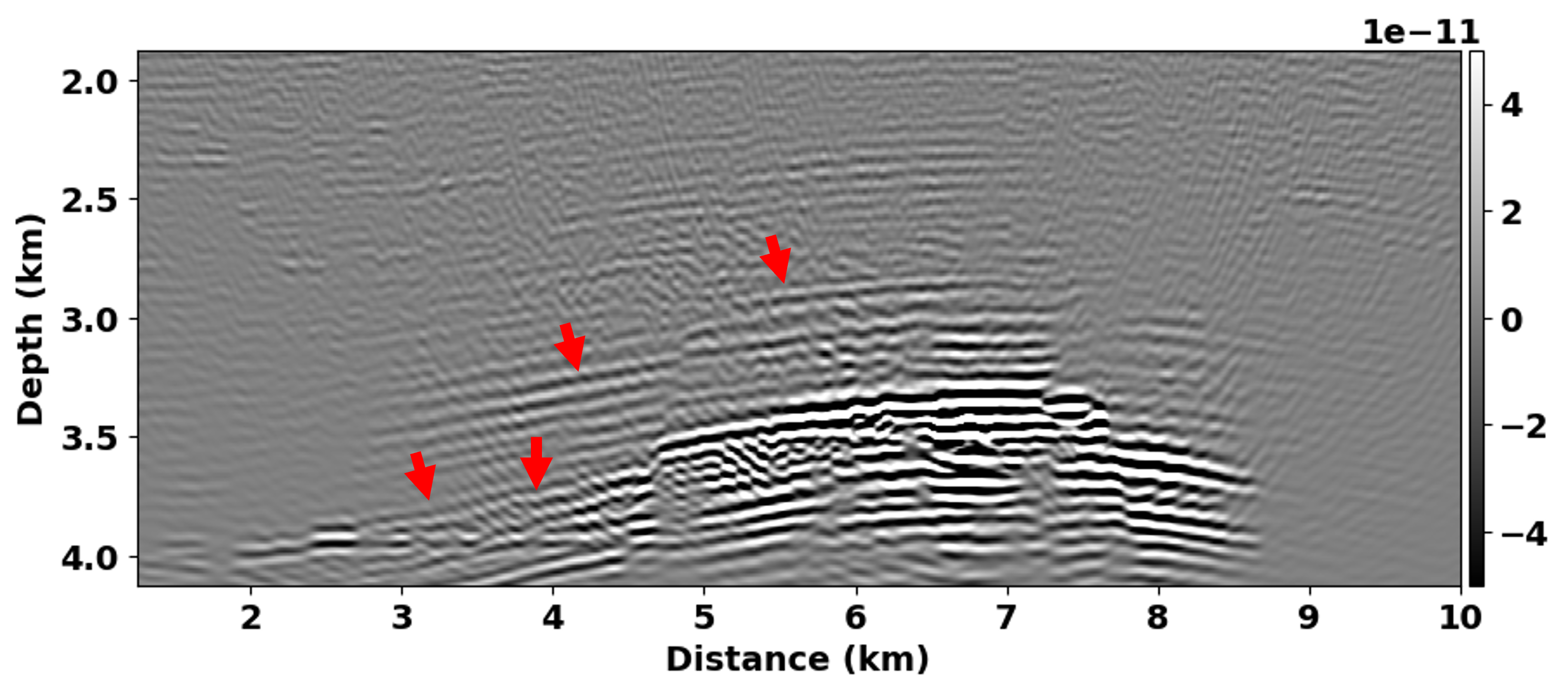}
    \label{fig:mf_img_seam}}
     \vspace{.1cm}
    \subfigure[]{\includegraphics[width=.6\textwidth]{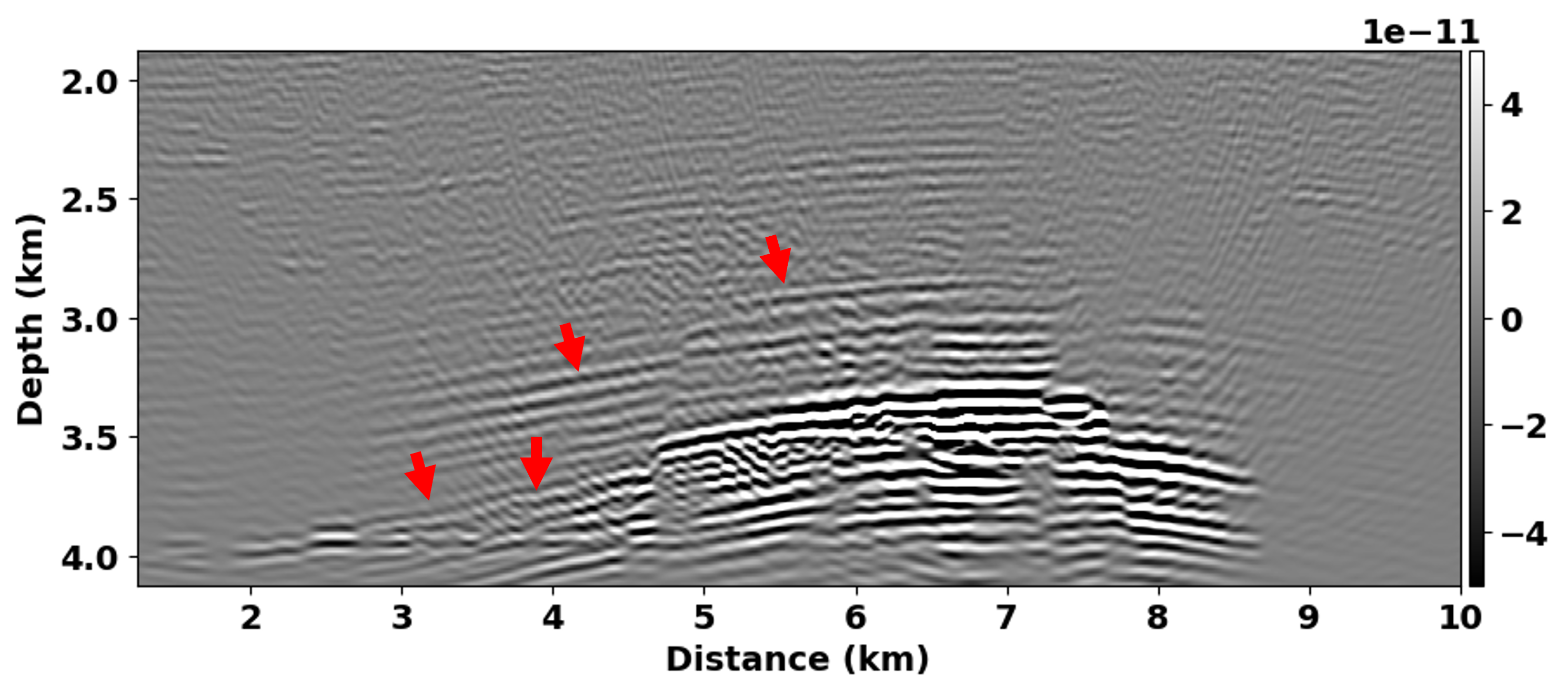}
    \label{fig:lstm_img_seam}}
    \subfigure[]{\includegraphics[width=.6\textwidth]{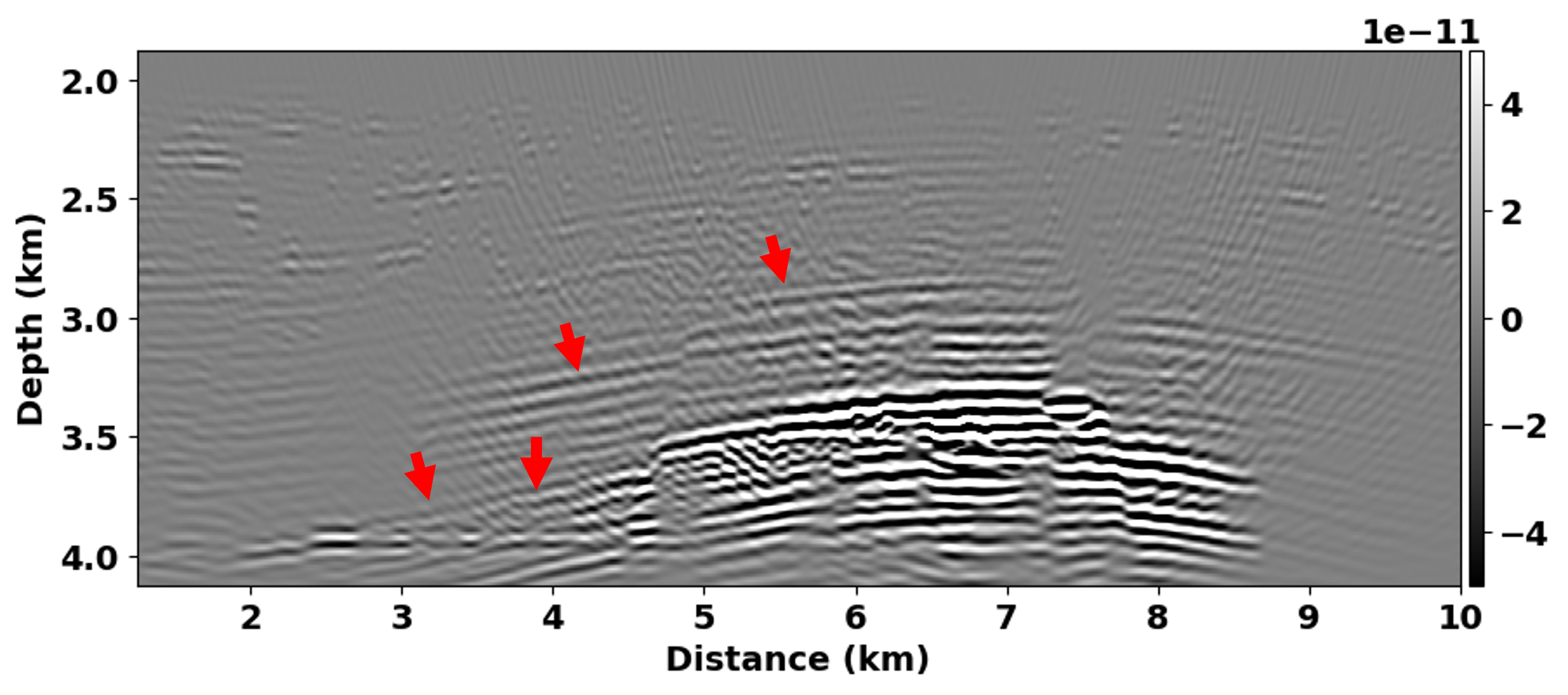}
    \label{fig:true_img_seam}}
    \caption{SEAM RTM images of the reservoir variation using the data differences between (a) the monitor and the baseline without processing, (b) after processing with the matching filter approach, (c) after processing with LSTM and (d) data corresponding to the true reservoir  variation. The red arrows point to some recovered 4D signal that was deformed by the near-surface variation.}
    \label{fig:img_seam}
\end{figure}

We also tested the method on the SEAM time-lapse model \citep{oristaglio2016seam}, given in Figure~\ref{fig:seam_base_model}. The reservoir changes, as shown in Figure~\ref{fig:seam_diff_model}, is now larger and more complex than before. The near-surface effects is designed similar to the previous example. We use 60 shots on the surface spaced by 175 m with a Ricker wavelet of 25 Hz dominant frequency to generate the data. We placed the receivers on the surface with a 25 m spacing. Figure~\ref{fig:Seam_shots} shows an example of a shot gather from the baseline, the target 4D signal and the difference between the baseline and the monitor, respectively. Examining the data difference (Figure~\ref{fig:seam_diff_shot}), we can identify the strong 4D reflections of the data that appear at around 3.2 s, but the weak 4D signals at 2.5 s are affected by the near-surface changes. 
\\
\\
In this example, we use two LSTM layers with a hidden size of 100, followed by a dense layer with tanh activation. We chose a training window from 1.3 to 2.1 s (350 time sample). We chose $w=40$ sample in the input trace, similar to the Otway example. The data consist of a total of 30060 traces. Similar to the previous example, we split the data such that 80\% of the traces used for training and 20\% for validation. The batch size is 64. We minimize the MSE loss for 300 epochs (Figure~\ref{fig:loss2}) using a learning rate of 0.002.
\\
\\
We then infer the data in the range 1.3-3.8 s. We plot five predicted shot in Figure~\ref{fig:pred_mon_seam}. Similar to the Otway example, we show the data differences before processing, after processing with the matching filter (filter-size$=w$), after processing with LSTM and the difference without the near-surface variation in Figure~\ref{fig:shots_diff_seam}(a), (b), (c) and (d), respectively. The NRMS and the predictability for the matching filter approach and the LSTM are very close to each other. 
\\
\\
Similar to the previous example, we use the data differences in Figure~\ref{fig:pred_mon_seam} to generate RTM images to better visualize the improvements in the reservoir region, as shown in Figure~\ref{fig:img_seam}. We can see many of the reflectors resulted from the overburden variations are suppressed and some other true reflectors are recovered as indicated by the red arrows. The SSIM value for the image before processing relative to the true image is 0.18, which is enhanced to 0.55 and 0.53 after processing with the matching filter and LSTM, respectively.

%% file: Texts/Noise.tex
\subsection*{Application on noisy data}
\begin{figure}
	\centering
	\subfigure[]{\includegraphics[scale=0.30]{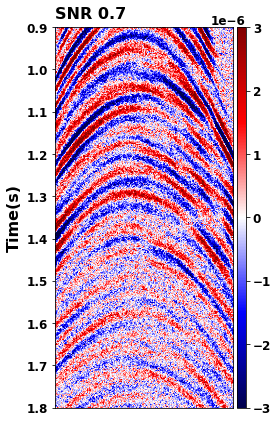}
		\label{fig:Otw_noise_mon7}}
	\subfigure[]{\includegraphics[scale=0.30]{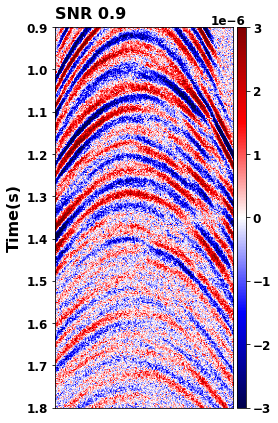}
		\label{fig:Otw_noise_mon9}}
	\subfigure[]{\includegraphics[scale=0.30]{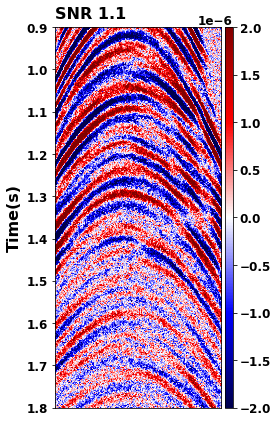}
		\label{fig:Otw_noise_mon1}} \\
		% -------------------------------------------------------------
	\subfigure[]{\includegraphics[scale=0.30]{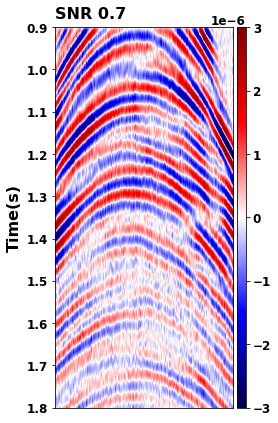}
		\label{fig:Otw_noise_pred7}}
	\subfigure[]{\includegraphics[scale=0.30]{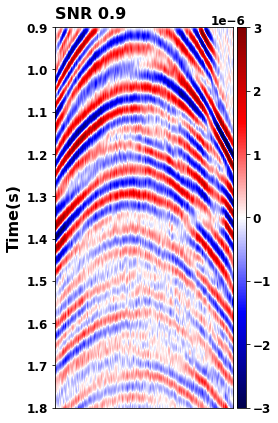}
		\label{fig:Otw_noise_pred9}}
	\subfigure[]{\includegraphics[scale=0.30]{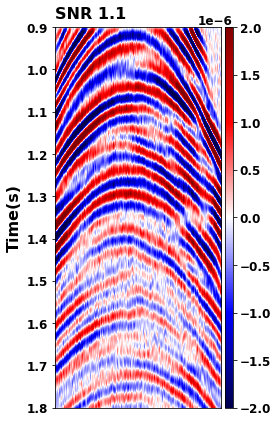}
		\label{fig:Otw_noise_pred1}}
	\caption{The first row are noisy shot gathers from the Otway model of the monitor data with SNRs 0.7 (a), 0.9 (b) and 1.1 (c). The second row are the corresponding predicted shot gather by LSTM. The predicted shot gathers appear to be cleaner and almost free of noise.}
    \label{fig:Otw_noise_shots}
\end{figure}

\begin{figure}
	\centering
	\subfigure[]{
	    \includegraphics[scale=0.30]{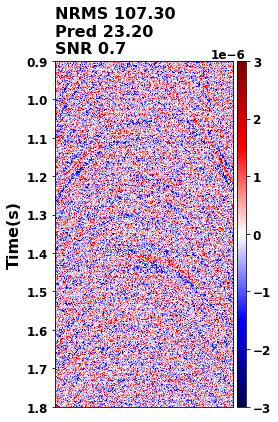}
	    \includegraphics[scale=0.30]{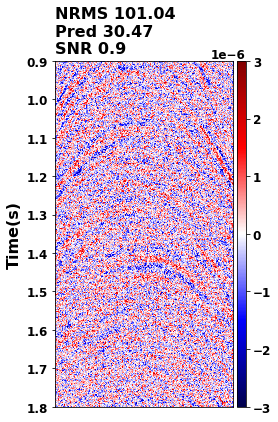}
	    \includegraphics[scale=0.30]{Fig/noisey_otway/before_diff_SNR0.9.png}
		\label{fig:Otw_noise_before}}
        \\
		% -------------------------------------------------------
	\subfigure[]{\includegraphics[scale=0.30]{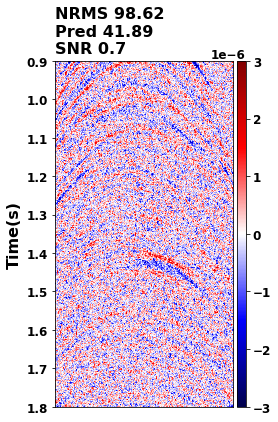}
                \includegraphics[scale=0.30]{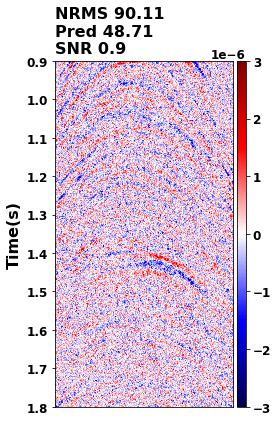}
                \includegraphics[scale=0.30]{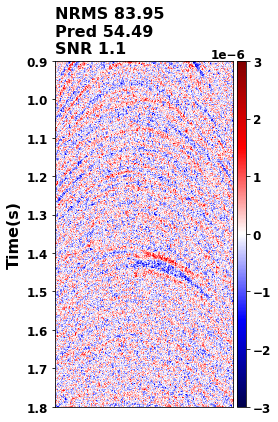}
		\label{fig:Otw_noise_mf}}
        \\
		% -------------------------------------------------------
	\subfigure[]{\includegraphics[scale=0.30]{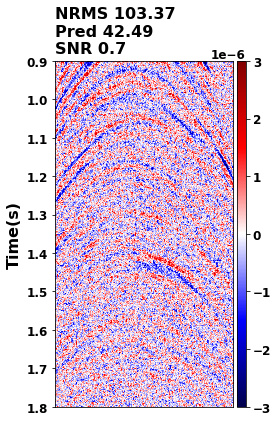}
            	\includegraphics[scale=0.30]{Fig/noisey_otway/before_diff_SNR0.9.png}
            	\includegraphics[scale=0.30]{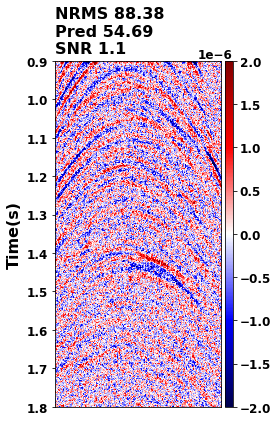}
		\label{fig:Otw_noise_lstm}}
        \\
		% -------------------------------------------------------
	\subfigure[]{\includegraphics[scale=0.30]{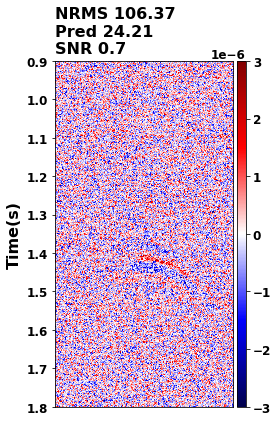}
            	\includegraphics[scale=0.30]{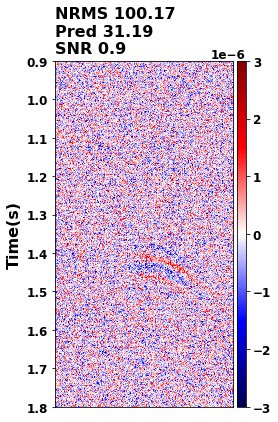}
            	\includegraphics[scale=0.30]{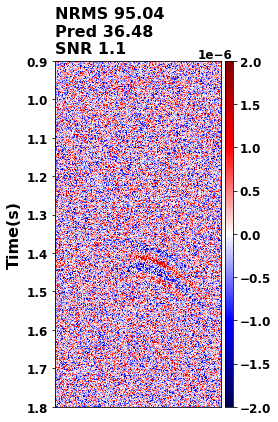}
		\label{fig:Otw_noise_target}}
	\caption{Data differences for the noisy data (a) before processing, (b) after processing with matching filter, (c) after processing with LSTM and (d) without the near-surface variations. Each column represents a certin SNR value as indicated on the top of the shots. NRMS and the predictability are sensetive to noise which result in poor values in all the shots.}
    \label{fig:Otw_noise_diff}
\end{figure}

In the previous examples, we showed the performance of the proposed approach on noise-free data. However, seismic data are often associated with various types of noise and testing the performance of the method on noisy data is essential. We add Gaussian random noise with different signal-to-noise ratios (SNR) to the base and monitor of the Otway data. We consider SNR values of 1.1, 0.9 and 0.7. Figures~\ref{fig:Otw_noise_shots}(a), (b) and (c) show one shot from the monitor for the three SNR values, respectively. We train the network similar to what we did in the clean data case and show the corresponfing predicted monitor data in Figures~\ref{fig:Otw_noise_shots}(d), (e), and (f). The predicted shot by the network is almost free of noise. In Figure~\ref{fig:Otw_noise_diff} we show the data differences before processing (Figure~\ref{fig:Otw_noise_before}), after processing with the matching filter (Figure~\ref{fig:Otw_noise_mf}) and with LSTM (Figure~\ref{fig:Otw_noise_lstm}. Examining the NRMS and the predictability metrics, we see that the LSTM and the matching filter are similar in their behavior. They both show a cleaner reflections from the reservoir variation, but they also include some unwanted residuals in the shallow part. We also plot the data difference without the near-surface variations, which barely show the reservoir variation. This is due to the fact that the variance of the random noise adds up when we subtract the data. However, when we take the difference in the LSTM case, which provides a noise-free prediction, we only struggle with the noise coming from the base.    

%% file: Texts/Discussion.tex
\section*{Discussion}
Our implementation is in general similar to the conventional matching filters. However, matching filters provide local matching capability and it does not take the evolution of the data with time into consideration. In contrast, using LSTM will take advantage of the time dependency in the data and learn the sequential behaviour of the data. In our tests, both approaches achieved somewhat similar results, mainly because our sources of time lapse variations (caused by the near surface) is stationary with time. We believe LSTM has the potential to overcome the conventional matching filter approach in cases where time-warping is needed before applying the matching filter for data alignment \citep{ayeni2009optimized}. Another case where learning the time-dependency might be useful is when there is an attenuation in the data. However, for the case of stationary in time, near surface variations, in the signal, LSTM performs as well as the matching filter approach.
\\
\\
A major assumption to this approach is that the characteristics of the data in the reservoir region is similar to that of the overburden. Thus, the differences between the base and the monitor is only coming from the shallow part and the reservoir imprints is small relative to the data and does not affect the distribution. However, if the data in the deeper part are reasonably different from the shallow part, the method is prone to fail.
\\
\\
The major hyper-parameters that we need to consider is the number of LSTM layers and the hidden layer size (neurons). Setting them to large values might lead to over-fitting, and small-size networks may not converge properly. Another important parameter in our approach is the $w$ in the input shape. In this research, these parameters were chosen based on trial and error and we chose the ones that give reasonable training-validation loss curves.
\\
\\
We only tested the approach for data under the acoustic constant- density assumption limiting the method to P-velocity changes. However, for elastic media the wavefield is more complex affected by the elastic properties of the medium such as the changes in the S-velocity and density. Including more elastic properties will add more complexity to the problem. Hence, it might require
increasing the learning parameters of RNN such as the hidden layer size or using a number
of stacked RNNs. We also suspect that in the elastic case, performing a 2D RNN or combining CNN with RNN such as in convolutional LSTM \citep{xingjian2015convlstm} will be helpful to capture the radiation patterns of the elastic properties.
\\
\\
The cost of this method highly depends on the training time, which depends on the size of the network (i.e. hidden-size and number of LSTM layer) and the data size. In our examples, we only show one monitor data where a full training was needed. If we have more than one monitor data, we can use transfer learning after training on the first monitor to adapt the network to the other monitor data, which will require fewer epochs to converge and lower cost.    

%% file: Texts/Conclusion.tex
\section*{Conclusions}
We proposed a new method to match the time-lapse data that utilized LSTM instead of the conventional matching filter. LSTM learns the time-dependency in the signal, which allows it to accommodate changes with time unlike the matching filter that is designed in a local window.
 We trained the network to map the data from the base to the monitor in the overburden area, and then inferred in the reservoir region. The difference between the predicted monitor and the actual monitor represent the 4D changes. 
 
 We provided three experiments to illustrate the potential of the method to match the data and compared them with the traditional matching filter approach. The first experiment is a simple matching between two traces, while the second and the third experiments are 2D synthetic data for the Otway and the SEAM time-lapse models, respectively. The repeatability for the data processed with LSTM is better than the data processed with the matching filter in the Otway example. However, in the SEAM example, the two approaches provided comparable results. In general, LSTM improved the repeatability and provided results that are comparable to those obtained by the matching filter. However, the examples used here only consider stationary time-lapse changes that do not evolve with time. In case of time-variant 4D changes exist, we believe the LSTM will outperform the matching filter courtesy of its ability to learn time-dependency. 
 
 RTM images corresponding to predicted data differences demonstrated how the LSTM can recover some deformed 4D signals. We further tested the approach on noisy data and found that the network enhanced the 4D reflections but it also provided unwanted residuals.

%% file: Texts/Appendix.tex
\section*{Appendix A}
\textbf{LSTM cell} \\

An LSTM unit consists of three gates: A Forget gate ($f_t$), that removes the useless information, an Input gate ($i_t$), which updates the internal states, and an Output gate ($o_t$), that forms the final output. The gates are composed of fully-connected neural networks combined with activation functions. Mathematically, LSTM updates the cell and hidden state as follow:  
\begin{align}
    f_t &= \sigma(W_f x_t + U_f h_{t-1} + b_f) \\
    i_t &= \sigma(W_i x_t + U_i h_{t-1} + b_i)\\
    o_t &= \sigma(W_o x_t + U_o h_{t-1} + b_o)\\
    g_t &= tanh (W_g x_t + U_g h_{t-1} + b_g)\\
    c_t &= f_t \bullet c_{t-1} + i_t \bullet g_t\\
    h_t &= o_t \bullet tanh(c_t), 
\end{align}
where {$g_t$} is the activation vector for the cell input, $W$ and $U$ are matrices for the weights and the recurrent connection, and $b$ is a bias vector. $\sigma$ represents the sigmoid activation function and ($\bullet$) is an element-wise multiplication.